\documentclass[manuscript=article,journal=jctcce,layout=twocolumn]{achemso} 
\usepackage[textsize=small]{todonotes}
\let\oldmaketitle\maketitle
\let\maketitle\relax           
\usepackage[utf8]{inputenc}
\usepackage{amsmath}
\usepackage{amssymb}
\usepackage{graphicx}
\usepackage{afterpage}
\usepackage{sidecap}
\newlength{\figonecol}
\usepackage[normalem]{ulem}

\newcommand{\da}[2]{{\ensuremath{{#1}_\text{D}{#2}_\text{A}}} }

\title{Probing defects and correlations in \\ the hydrogen-bond network of ab initio water}

\author{Piero Gasparotto}
\affiliation{Laboratory of Computational Science and Modeling, IMX, {\'E}cole Polytechnique F{\'e}d{\'e}rale de Lausanne, 1015 Lausanne, Switzerland}

\author{\\Ali A. Hassanali}
\email{ahassana@ictp.it}
\affiliation{The Abdus Salaam International Center for Theoretical Physics, Condensed Matter Physics Section, Strada Costiera 11, Trieste Italy.}

\author{Michele Ceriotti}
\email{michele.ceriotti@epfl.ch}
\affiliation{Laboratory of Computational Science and Modeling, IMX, {\'E}cole Polytechnique F{\'e}d{\'e}rale de Lausanne, 1015 Lausanne, Switzerland}

\date{}
  \newlength{\stdwidth}
\begin{document}

\twocolumn[
\begin{@twocolumnfalse}
\oldmaketitle
\begin{abstract}
The hydrogen-bond network of water is characterized by the presence of 
coordination defects relative to the ideal tetrahedral network of ice,
whose fluctuations determine the static and time-dependent
properties of the liquid. Because of topological constraints, such defects 
do not come alone, but are highly correlated coming in a plethora
of different pairs. Here we discuss in detail
such correlations in the case of \emph{ab initio} water models and
show that they have interesting similarities to regular and defective
solid phases of water. Although defect correlations involve deviations from
idealized tetrahedrality, they can still be regarded as weaker hydrogen bonds 
that retain a high degree of directionality. We also investigate how 
the structure and population of coordination defects is affected by 
approximations to the inter-atomic potential, finding that in most cases,
the qualitative features of the hydrogen bond network are remarkably robust.

\end{abstract}
\end{@twocolumnfalse}
]


\section{Introduction}

Many of the anomalous physical and chemical properties of water can
be understood in terms of its highly-structured hydrogen-bond network
\cite{Bergman2000,AgmonReview2011}. 
Tetrahedrally coordinated water, with two donated and two
accepted H-bonds constitutes the fundamental building block of such 
networks.  Of course, this idealized tetrahedral environment 
can be heavily distorted by thermal\cite{KuhneKhalliulin2013} and 
quantum\cite{CeriottiCunyParrinelloManolopoulos2013} fluctuations. 
In the liquid phase, coordination defects exist and their presence,
concentration, and relative arrangement contribute 
to the structural and dynamical properties of
water\cite{SciortinoFornili1989,henc-irud10jpcb,AgmonReview2011}. 
Here we investigate
the properties of such coordination defects, with a particular
focus on their structural correlations, by means of first-principles
molecular dynamics simulations. 

Over the last three decades, considerable effort and progress has been
made in the simulation of liquid water from first principles
calculations. In this regard, DFT-based ab initio simulations
have elucidated the importance of several factors such as
the quality of the electronic structure, the
treatment of nuclear quantum effects (NQE) and also the role of
statistical sampling
\cite{KirchnerDioHutter2012,Sprik1996,
  Todora2006,GuidonSchiffmanHutterVandeVondele2008,
  SilvestrellliLuigiParrinello1999,
  GrossmanSchweglerDraegerGygiGalli2003,SerraArtacho2004,SitMarzari2005,
  VandeVondeleMohamedKrackHutterSprikParrinello2005,LeeTuckerman2006b,
  KuoMundyMcGrathSiepmannVandeVondeleSprikHutterKleinMohamedKrackParrinello2004,
  Zhang2011,LinSeitsonenCoutinhoMauricioTavernelliRothlisberger2009,
  SantraMichaelidesScheffler2009,JohnchiereSeitsonenFerlatSaittaVuilleumier2011,CuiWuGalliGygi2011,
  AndreasKelkkanenAndreWikfeldtJakobJorgenLArsJacobsenNillssonJens2011,
  Chen2003,MorroneCar2008,delb+13jpcl,DiStasioBiswajitZhaofengXifanCar2014}
in reproducing the experimentally available 
oxygen-oxygen pair correlation function
of water (RDF see Figure~\ref{fig:goo-ref}). Unlike
the idealized tetrahedral structure of ice, finite temperature
fluctuations create defects which are a small fraction
of the H-bond network and thus challenging to sample.
Here we have taken exceptional 
precautions to ensure as extensive as possible thermodynamic sampling
to collect meaningful statistics on the populations and structure
of defects that in some cases contribute to less than a percent
of the H-bond network of liquid water. To achieve this, we used
parallel tempering combined with the well-tempered ensemble\cite{bono-parr10prl}
(PTWTE) to perform an extensive sampling of a box of 64 and 128 water molecules. 

Using different models of inter-atomic forces, or different thermodynamic
conditions, may change slightly the typical geometry of a hydrogen bond. 
In order to make the structural definition of a hydrogen bond and 
the classification of the different coordination environments 
independent from these effects, we used an adaptive, probabilistic 
definition of the hydrogen bond (PAMM)\cite{gasp-ceri14jcp}.
Since in this work we restrict ourselves to a narrow range of 
thermodynamic conditions, this choice does not entail dramatic 
differences relative to one of the more traditional definitions, 
but provides a robust framework that would make it
straightforward to perform a similar analysis for different 
systems or to investigate more dramatic changes in environmental factors. We also report on the impact of different 
details of the electronic structure calculation,
although as we will see,
for a given choice of exchange-correlation functional, the main 
factor contributing significantly to the structure 
and population of defects is the use or neglect of dispersion corrections. 
This is consistent with previous observations which have shown
that the inclusion of dispersion corrections significantly 
reduces the over-structuring that is 
seen for the most common choices of 
exchange-correlation density functional
\cite{LinSeitsonenCoutinhoMauricioTavernelliRothlisberger2009,YooXantheas2011,AkinOjoWang2011}

The topological constraints  induced by the presence of
an extended H-bond network mean that defects appear at 
highly correlated positions and are hence clustered together
with different propensities~\cite{StanleyTeixeira1980,henc-irud10jpcb}. 
What is more, the details of the description of the 
inter-atomic forces do not significantly change 
the population of defects, nor their relative
structural correlations. 
Although the RDF provides useful information on the structure of
the system that can be readily compared with accurate experiments\cite{Soper1986} 
and is therefore regarded as the holy grail for benchmarking the quality  of
ab initio models, it averages over 
all the underlying complexity of the
topology of the HB network and its
directional correlations\cite{HassanaliGibertiCunyKuhneParrinello2013}.
As we will show here, one could consider the RDF
as arising from the combination of correlations
between different ideal or defective coordination environments.
We present an extremely thorough analysis of such correlations, 
assessing the impact of many different computational details,
and showing which of those matter, and which only cause small changes
to the RDF but no profound qualitative change to the topological
properties of the H-bond network.
Furthermore, we use defect-resolved three dimensional distribution
functions to elucidate the role of the weak interactions that 
are characteristic of under-coordinated environments.
We find that  the interactions between defects
formed in the network have a remarkably directional character that
could be interpreted as arising from `weak' rather than altogether
`broken' hydrogen bonds -- and link these angular 
correlations to those found in solid phases of water. 
While our analysis here focuses on thermodynamic, time-independent
properties, we believe the structure and correlations of H-bond
defects will prove crucial to understand the fluctuations
and hence dynamics of liquid water in future studies. 

The paper is laid out as follows: we begin by describing the systems
we simulated and the computational methods we used in
Section~\ref{sec:CompMethods}.  
In Section~\ref{sec:defectpop} we
catalogue the structural patterns in ab initio water obtained with
the PAMM analysis. In
Section~\ref{sec:defectcorr}  we use the patterns detected by PAMM
to infer structural correlations between defects in water's hydrogen bond
network. In Section~\ref{sec:watermodelcomp} we compare
the distribution of coordination defects produced by
different water models varying the level of theory used
for the electronic structure, finite box size and nuclear
quantum effects. Finally we end with some concluding remarks on our work in
Section~\ref{sec:Discussion}.

\section{Computational Methods}
\label{sec:CompMethods}

In this work we have used AIMD simulations based on density functional
theory (DFT) coupled with PTWTE.
We will begin by first summarizing details of the electronic
structure methods used and then describe the protocol we applied for the
PTWTE simulations. 

\subsection{Ab Initio methods}

The electronic structure calculations for computing the energies and
forces were conducted using Quickstep which is part of the CP2K
package\cite{VandeVondeleKrackMohamedParrinelloChassaingHutter2005}.
The molecular dynamics and parallel tempering simulations were performed
using the recently released code i-PI, that decouples the calculation
of the interatomic forces from the dynamic evolution of the
nuclei\cite{ceri+14cpc}. A convergence criterion of
$5\times10^{-7}$a.u  was used for the optimization of the 
wavefunction in all the simulations. Unless otherwise stated, the 
wavefunction was expanded  in a DZVP Gaussian basis set, and an auxiliary
basis set of plane waves  was used to expand the electron density up
to a cutoff of 300Ry. The D3 Grimme dispersion corrections
\cite{grim+10jcp} for the van der Waals interactions were used for most of the
simulations. We used the BLYP generalized gradient
correction\cite{LeeYangParr1988} to the local density approximation
and Goedecker-Teter-Hutter (GTH) pseudopotentials~\cite{GoedeckerTeterHutter1996}. All simulations were 
thermostated within the NVT ensemble
using the canonical-sampling velocity-rescaling
thermostat\cite{buss+07jcp} with a time constant of 
1fs. To maximize sampling of uncorrelated
potential energy structures and accelerate replica exchanges 
in PT simulations,
we also included a generalized Langevin equation thermostat tuned
for efficient sampling~\cite{ceri+09prl,ceri+10jctc}.  

Extensive tests
of the sensitivity of results to different computational details were 
performed using a box of side length 12.4138\AA{} with 64 water
molecules, corresponding to the experimental density of the system at 300K. 
The different PT simulations that were conducted included the
following: BLYP without D3 Grimme's
dispersion correction (BLYP+NOVDW), BLYP
with dispersion corrections (BLYP+VDW), 
BLYP+VDW simulations with the TZV2P basis set (BLYP+VDW+TZV2P) and
finally BLYP+VDW simulations using a plane wave
cutoff of 350Ry (BLYP+VDW+350). For all the previously described
PT runs, a timestep of 1fs was used. 
Most AIMD simulations using Born
Oppenheimer molecular dynamics use a smaller timestep of
0.5 fs, which is necessary to obtain accurate real-time 
dynamics, but does not change significantly structural properties. 
In order to assess the sensitivity of our results to the choice
of a larger timestep than commonly used, PT simulations were  also
conducted using BLYP+VDW with a timestep of 0.5fs (BLYP+VDW+0.5fs)
which shows that using a larger timestep does not qualitatively change
the structural properties of the system such as the
diversity of different defects in the system (for more details
see Figure. S3 of the SI).

The PT runs for the 64 water boxes detailed above were used to
identify parameters for our production simulations with a larger box.
We thus also performed parallel tempering simulations
of 128 water molecules with a box size of 15.6404\AA{} 
using BLYP+VDW, DZVP, 300Ry cutoff and 1fs time step. This
simulation will be referred to as PTL.
From our PTL simulations, we initiated 
four independent PIGLET simulations, using six beads and 
colored-noise thermostatting\cite{ceri-mano12prl}, 
to assess the role of nuclear quantum effects (NQE).

Besides the simulations using the BLYP functional, we also conducted
simulations with the more expensive hybrid functional B3LYP implemented
in CP2K\cite{guid+10jctc}, 
with D3 vdW corrections,  in
order to ascertain the role of electron exchange on the properties
of liquid water. Due to the high cost of including Hartree-Fock
exchange, we could not run replica-exchange simulations. Instead, we 
ran four independent simulations of about 16ps each, starting
from BLYP equilibrated configurations extracted from the BLYP+VDW PT
simulation consisting of 64 waters, and 
discarding the first 2ps for equilibration with the hybrid
functional.

Although the combination of hybrid functionals and dispersion
corrections appear to improve the properties of
ab initio water, we wanted to ascertain whether the
defect correlations were not exclusive to
DFT simulations. Hence, to complement our AIMD simulations,
we also investigated populations and correlations of 
defects using force field models. Namely, we performed
a classical simulation of 512 water molecules using a fixed point-charge 
model (TIP4P/2005f \cite{gonz-abas11jcp}) and 
classical and PIGLET simulations of 216 water 
molecules using MB-pol,
a sophisticated force-field based on a many-body
expansion and high-end quantum chemical reference 
calculations\cite{MBPOL1,MBPOL2,MBPOL3}. 
For non-reactive simulations of water, MB-Pol
is probably the most realistic theoretical model of the 
behaviour of molecular H$_2$O. 
Finally, we also make some comparisons of the topological defects
found in liquid water to the coordination environment found in 
solid phases of ice.

In what follows, we will focus mostly on the results of the PTL
simulations with 128 waters and make reference to the analysis
of the temperature dependence and electronic structure in the text,
but deferring most of the details to the SI. Other simulation details,
including the lengths of all the simulations conducted,
 are summarized in Table I of the SI.

\subsection{Parallel tempering protocol}

Ab initio parallel tempering simulations for the 64 water boxes 
were performed using  six
replicas at the following temperatures: 290, 304, 322, 343,  365, 390K. 
In parallel tempering, a series of replicas 
are simulated at these six temperatures and thereafter
exchanges between adjacent replicas are performed using a Metropolis
criterion. In order to achieve efficient exchange between the replicas
there must be significant overlap in the potential energy
distributions of neighboring temperatures. Since a larger system 
exhibits smaller (relative) energy fluctuations, for PTL we used
8 replicas at the temperatures 290, 300, 310, 322, 335, 351, 369, 390K.
It has recently been shown
that by combining parallel tempering with the well tempered ensemble
(PT+WTE), the overlap in energy between adjacent replicas is increased
resulting in more frequent exchanges\cite{bono-parr10prl}. In the well 
tempered ensemble the enhancement of the fluctuations in energy is 
tuned by the $\gamma$ factor which in our case was chosen to be 4. Exchanges between the
replicas were attempted every MD step. 

AIMD simulations are still
prohibitively computationally expensive and hence it is rather challenging
to perform a systematic benchmarking of parameters to tune the optimal
number of replicas used as well as the $\gamma$ factor. We decided to adopt a simplified
protocol to determine the bias to generate the well-tempered ensemble.
The mean and fluctuations of potential energy as a function of temperature
were determined based on a short (5ps) preliminary run. Then, a fixed 
bias was constructed and applied to the remainder of the simulation. Rather than using 
a (divergent) parabolic bias, we decided to give it a Gaussian shape with mean 
and curvature compatible with the
measured fluctuations, but designed to cut off to zero for large fluctuations. Specifically,
we used
\begin{equation}
 B(V)=k_B T \left(\gamma - 1\right) \exp \left[-\frac{1}{2 \gamma \delta V^2} \left(V - \bar{V}\right)^2\right],
\end{equation}
where $\bar{V}$ and $\delta V^2$ are respectively the mean and variance of the
potential energy evaluated for a given parallel tempering replica. 

\begin{figure*}[tbp]
\includegraphics[width=0.9\textwidth]{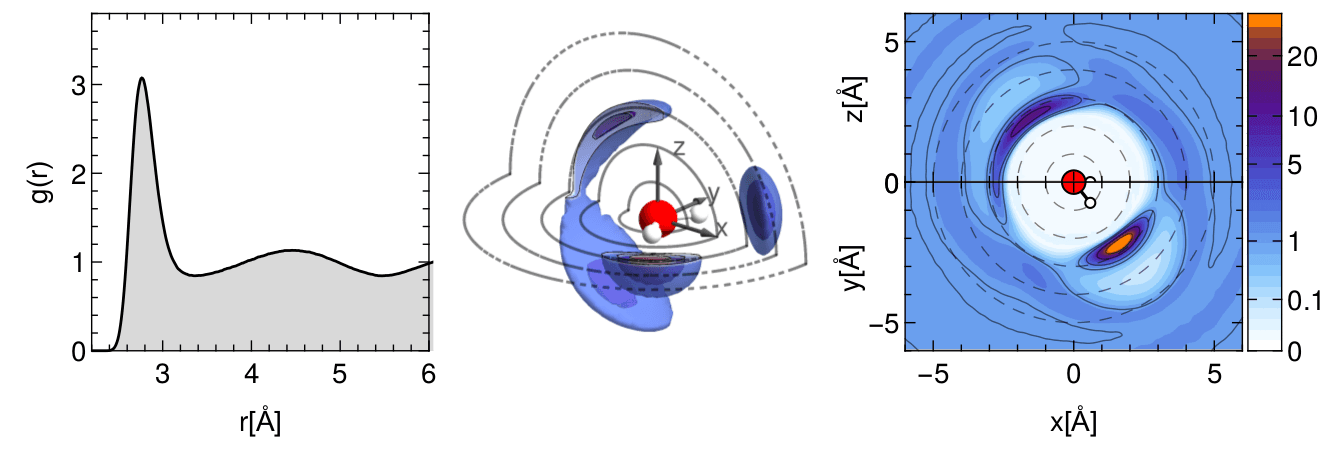}
\caption{\label{fig:goo-ref}Oxygen-oxygen radial distribution functions computed at 300K from PTL (BLYP+VDW+PTWTE-128) runs. 
Left to right, the figures correspond to the usual 1D radial distribution, 
to the 3D distribution and finally 
to slices of the 3D distribution along the $xy$ and $xz$ 
molecular planes. Dashed circles are a drawn as a guide for the eye, indicating 
a radial grid with a 1\AA{} spacing.}
\end{figure*}

\section{Structural patterns in ab initio water}
\label{sec:structpatterns}

Let us start by introducing the different types of pair correlation
functions that will be discussed in the rest of the paper. 
The left-most panel of Fig.~\ref{fig:goo-ref} illustrates the
familiar oxygen-oxygen pair RDF obtained from the PTL simulations at 300K. 
The middle panel of Fig.~\ref{fig:goo-ref} shows instead the 3D 
oxygen-oxygen correlation function, which contains information
on angular correlations that are lost in the RDF. The contour plot
in the right-most panel shows a cut of such a 3D correlation function
in the plane of the molecule ($xy$) and along the orthogonal symmetry plane ($xz$). 
The middle and right panels quantify the orientation of water
molecules either accepting or donating hydrogen bonds in
the first hydration shell of a particular water.
The peaks at positive $x$  correspond
to the position of water molecules to which the tagged water is
donating a hydrogen bond, and the broader peaks at negative
$x$ values involve the accepting side of the  central molecule.
This is consistent with the notion of a ring of delocalized electron
density, refered to as the 'negativity track' created by the 
lone-pairs which permits a larger range of local 
tetrahedrality\cite{AgmonReview2011}. Note
that a peak in the density from both the donating and accepting
side is a signature
of highly directed nature of the hydrogen bonds. The importance
of highlighting this feature will become clear when we show
similar distributions for clustered defects. 

One of the crucial observations we make is that
there are significant correlations amongst
defects in the hydrogen bond network. Since these defects tend
to be rare and fleetingly lived and characterized by
geometrical properties that are possibly different from
idealized tetrahedral water, we wanted to ascertain how
structurally intact or well-defined hydrogen bonds change under
different definitions or thermodynamic conditions. 
In a recent study, it was proposed that an unbiased definition of
general fingerprints functions for recurring molecular patterns could 
be obtained automatically from an analysis of simulation data.
Such a probabilistic analysis of molecular motifs (PAMM)~\cite{gasp-ceri14jcp}
was shown to give an adaptive definition of H bonds, that could 
be made fully consistent with systems as diverse as alanine dipeptide,
classical and quantum water, and liquid ammonia. The reader
is referred to the original paper for details of the method, which
is based on the automatic identification of local maxima in
the probability distribution in pattern configuration space, 
followed by optimization of a Gaussian mixture model. The posterior
probabilities give a natural partitioning of the configuration
space over stable, recurring molecular patterns.

This definition has many advantages over more traditional ones.
Firstly, it is probabilistic in nature and
fuzzy: each O--H$\cdots$ O$'$ triplet is assigned a fingerprint that varies 
smoothly between 0 (no HB) and 1 (clear-cut HB). Secondly, it is adaptive:
since it detects modes of the probability distribution in configuration
space, the definition will change depending on temperature and water
model, separating clearly the slight model dependence of HB geometry and
the changes in populations of defects. Whereas a conventional
geometrical definition would require a manual adjustment of its 
parameters~\cite{kuma+07jcp}, the PAMM algorithm
determines automatically, for each simulation scenario, 
which range of geometries should
be considered as a hydrogen bond. For the simulations conducted in this 
specific study -- which is  performed at the thermodynamic conditions to 
which the traditional
hydrogen-bond definitions have been tuned -- the choice of 
a PAMM definition over a conventional bond-angle definition introduces
only minor differences and does not change our 
conclusions (see Section.3 in SI).

\begin{figure}[btp]
\includegraphics[width=1.0\columnwidth]{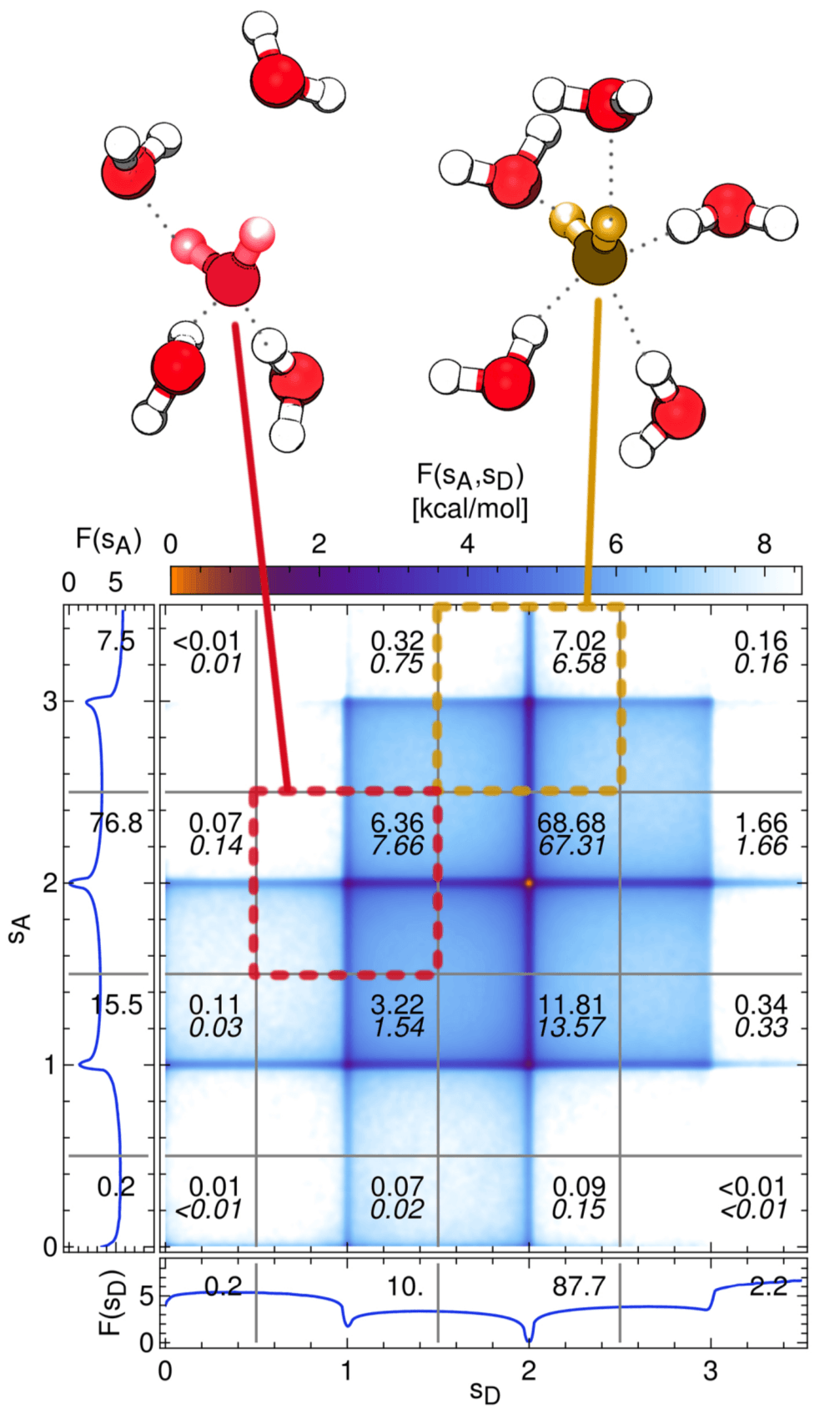}
\caption{\label{fig:hb-fingerprint} H-bond coordination summary
for the PTL simulation data at $T=300$K. $s_\text{D}$ counts
the H-bonds donated by a O atom, and $s_\text{A}$ those
accepted. Fractional values characterize fluctuations. 
The $(s_\text{D},s_\text{A})$ range is partitioned in discrete
regions that are assigned to different coordination states. 
For instance, the region with $0.5\le s_\text{D} < 1.5$ and $1.5\le s_\text{A}<2.5$
is assigned to the $\da{1}{2}$ state. The numbers reported
in each region correspond to the overall fraction of O atoms 
observed in a given state (in percent), while the number in italics
is the corresponding value obtained as a product of the marginal 
probabilities. The larger the difference between the two values, 
the larger the correlations that exist between the donor and acceptor 
counts, for a given coordination state. 
}
\end{figure}

\subsection{Population of coordination defects 
\label{sec:defectpop}}

The PAMM methodology provides a convenient approach to
define three main types of H-bond count functions, $s_\text{D}$, $s_\text{A}$, 
and $s_\text{H}$. $s_\text{D}$ quantifies the total number of HBs donated by a
tagged O atom, $s_\text{A}$ the number of HBs accepted by an atom and finally
$s_\text{H}$ quantifies the number of HBs that any particular
hydrogen participates in~\cite{gasp-ceri14jcp}. These are obtained by 
summing the value of the PAMM HB fingerprint computed for 
all possible donor-H-acceptor triplets involving the tagged
atom\footnote{Of course, in practical implementations, a cutoff and
a neighbor list can be used to maintain a linear scaling computational complexity
of the analysis.}.  Based on these counts, one can build very informative defect stability
maps as seen in Fig.~\ref{fig:hb-fingerprint}, 
that summarizes the relative probabilities of 
finding a water molecule in one of different coordination states. It is clear
that for the PTL simulations at 300K, most water molecules
donate and accept 2 hydrogen bonds. However, there is a sizable
fraction of different types of topological defects. 
A nice feature of this type of analysis is that one can immediately point to
asymmetries in the accepting vs donating abilities of hydrogen bonds - 
for example, there is a higher probability of finding water molecules
that accept 2 and donate 1 hydrogen bond compared to those that
accept 1 and donate 2 hydrogen bonds. We also see clearly 
the asymmetry in the distributions associated with
a water molecule being a donor or acceptor which is consistent with
previous observations by Agmon\cite{AgmonReview2011}. Similar maps
for other simulations at a higher temperature and electronic
structure approximations are shown in the SI.

\afterpage{
\begin{SCfigure*}[1][!p]
\caption{The figure reports concisely the O-O correlation functions involving
a defective H-bond O environment, and a second oxygen
atom without specification of its H-bonding state.
The first row reports the baseline O-O correlations,
as in Fig.\ref{fig:goo-ref}. The 1D radial distribution
function reports all the distributions of oxygen atoms
around the most important defect environments, with 
the baseline shaded in grey. 
The following six panels report slices along 
high-symmetry molecular planes of the 3D distribution
of the various defects around a non-specified O atom,
while the last six report the mirror distribution of
an arbitrary O atom around the specified defective 
species. \label{fig:OO-all}\\
\protect\rule{0ex}{5cm}}
\includegraphics[width=0.75\textwidth]{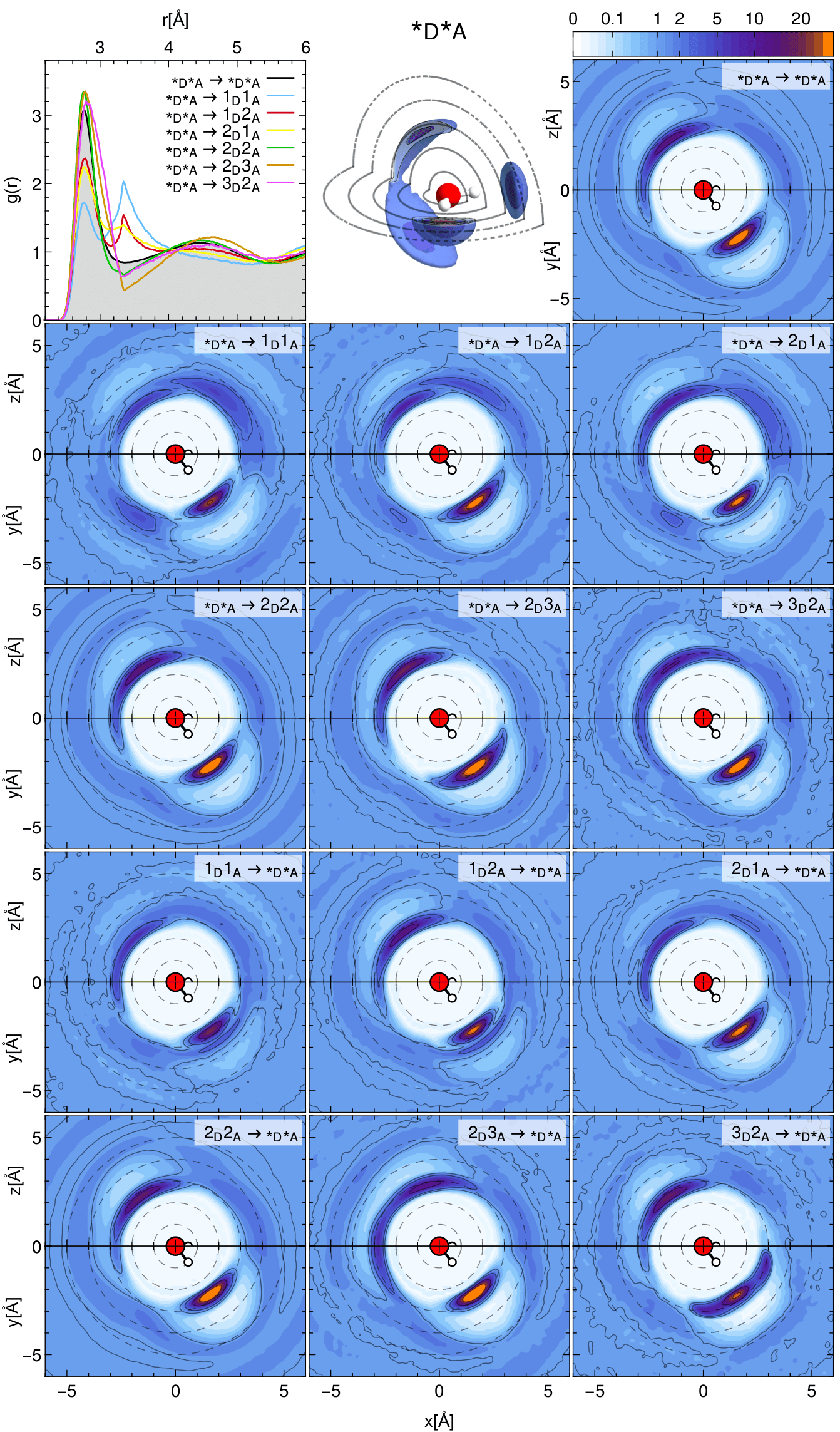}
\end{SCfigure*}
}

An important advantage of PAMM relative to traditional H-bond definitions
is that since the underlying fingerprint is probabilistic, and
varies smoothly between zero and one, it is possible to
recognize features
in the transition regions between clear-cut defect states.
The smooth definition of PAMM provides a qualitative description of 
the pathway from one defect state to another.
Even though an order parameter based on a single site cannot 
capture quantitatively the complex collective modes that underlie
the rearrangement of the H-bond network~\cite{LaageHynes2008},
the height of the barrier between two defect states gives an 
indication of the relative propensity towards a transition. 
For instance, one can see that in the overwhelming majority of
cases the state of an O atom evolves by increasing or decreasing by 
one the donated or accepted hydrogen-bond count, while concerted transitions
do not contribute significantly. 
By looking at cuts in the free energy surface at constant
$s_\text{A}$ or $s_\text{D}$, one can also get an idea
of how the free energy barriers along the pathways to make
or break hydrogen bonds change for molecules that are initially
undercoordinated or overcoordinated (see Figure S1 in the SI).
For instance, the barrier for the $\da{1}{2}\rightarrow\da{1}{1}$ transition
is lower by $\sim$ 33$\%$  than the barrier for the $\da{2}{2}\rightarrow\da{2}{1}$
transition. Similarly, the barrier for the $\da{2}{1}\rightarrow\da{1}{1}$ 
transition is lower than the $\da{2}{3}\rightarrow\da{1}{3}$ 
transition roughly by a factor of two. Interestingly, these
qualitative trends appear to be quite robust to the choice
of different approximations made in treating the underlying
electronic potential. However, it should
be noted that small differences in these barriers
will lead to much larger changes in dynamical properties,
thus making it even more challenging to achieve statistical
convergence.

Although we will not exploit this aspect here, it is worth
stressing that strictly speaking, PAMM identifies coordination
of O atoms, rather than water molecules, which means that this
classification could be used transparently also in the presence
of charged defects and charge fluctuations, 
since one is not relying on
the definition of molecular entities. 
For instance, in the presence of an excess proton, 
one would expect to see an increase in density 
in the region with $s_\text{D}\approx 3$ and 
$s_\text{A}<1$, or detect the quantum fluctuations
of a proton along a hydrogen bond by the appearance
of diagonal features (see Ref.~\cite{gasp-ceri14jcp}) 
that correspond to one hydrogen bond momentarily 
changing its character from acceptor to donor.

\subsection{Defect correlations and the RDF}
\label{sec:defectcorr}

The hydrogen bond maps of $s_{A}$ and $s_{D}$ shown in 
Fig.~\ref{fig:hb-fingerprint} provide a convenient way to classify
water environments based on their coordination state~\cite{lawr-skin03cpl,auer+07pnas}.  
Given that the probability
maxima are very clear-cut and with a rather obvious structure as seen
in Fig.~\ref{fig:hb-fingerprint}, we subdivided
the map manually labelling e.g. $\da{n}{m}$
an oxygen atom that has
$s_\text{A}\in \left[m_\text{A}-0.5,m_\text{A}+0.5\right)$ and  
$s_\text{D}\in \left[n_\text{D}-0.5,n_\text{D}+0.5\right)$. 

Armed with this classification, we can proceed to investigate
whether different $\da{n}{m}$ states are distributed randomly in
the network, or whether significant correlations exist between them.
Two-body spatial correlation functions provide powerful tools
to recognize such correlations, and allow us to disentangle the underlying factors that control
the shape of the overall O-O RDF shown in Fig.~\ref{fig:goo-ref}.

The simplest analysis involves coordination-resolved RDFs $\da{n}{m}-\da{n'}{m'}$ 
-- that report on the probability of finding an O atom in a coordination
state $\da{n}{m}$ and another in $\da{n'}{m'}$ at a distance $r$ from one 
another. In addition, one can also look at the 3D distributions, as
shown in Fig.~\ref{fig:goo-ref}, which give deeper insight
into the angular position of waters within the first hydration shell.
We will label as $\da{n}{m}\rightarrow \da{n'}{m'}$ 
the distribution of $\da{n'}{m'}$ evaluated in the reference frame of a 
$\da{n}{m}$ molecule. It is important to recognize that for each pair of species, the 
distributions associated with $\da{n}{m}$ and $\da{n'}{m'}$ are
\emph{not} symmetric. If there is enhanced probability of finding
$\da{n'}{m'}$ waters in the donor region of a $\da{n}{m}$ molecule,
one can expect that viewed from the point of view of 
$\da{n'}{m'}$ this same correlation will amount to an
enhancement in the \emph{acceptor} region.
Figure~\ref{fig:OO-all} provides an example of this kind of
analysis, where we consider correlations between an un-specified $\da{\star}{\star}$ O atom
and the main coordination states.
The complete series of radial and 
3D correlation functions, for all temperatures 
and electronic structure methods we considered in the present study, are provided in the
SI. Here we will only comment on the most significant correlations, that
can shed some light on the complex topological features of the H-bond network
of liquid water.

Let us start by commenting on the O-O radial distribution functions.
It is tempting to analyze the changes of the overall O-O RDF 
with temperature and simulation details in terms of the components resolved 
into the different coordination states. Contrary to the inherent structure
analysis, that identifies ``low-density/high-tetrahedrality'' 
and ``high-density/low-tetrahedrality'' by quenching instantaneous liquid
configurations\cite{WikfeldtNilssonPettersson2011}, the
analysis we perform here includes snapshots that
are fully consistent with the given thermodynamic state point. 
We have seen in Section~\ref{sec:defectpop} that the majority of water molecules 
corresponds to thermal fluctuations around a tetrahedral $\da{2}{2}$ environment. 
All defective environments conspire to modify the shape of the
short-range region of the RDF: over-coordinated defects do so by broadening the first
peak, but would themselves increase the depth of the first minimum. 
Undercoordinated environments, instead, enhance the density in the interstitial
region. This analysis thus provides an alternative interpretation of the
structure of the RDF in terms of correlations between coordination defects
on top of a dominant tetrahedral network, without invoking 
explicitly the existence of two thermodynamically 
distinct low and high-density water networks. 
As can be seen in the SI, changing the 
simulation temperature, or the details of the
electronic structure, modulate both
the populations and the shape of the RDF of defect resolved individual components, 
although many of the qualitative features associated with the relative
population of defects as well as the strutural correlations between them,
are still conserved (see SI for details).

\begin{figure*}[hbt]
\caption{Defect-resolved structural correlations in the 
hydrogen-bond network of vdW-corrected BLYP water. We focus on a
few particularly significant pairs of under-coordinated
and over-coordinated defects. As described in the main text, 
the structural correlations resemble some features seen
in solid phases of ice. The crosses and circles correspond, respectively,
to the positions of the nearest oxygen atoms in ice Ih, and
ice VIII expanded to match the density of room temperature water.
}
\includegraphics[width=0.75\textwidth]{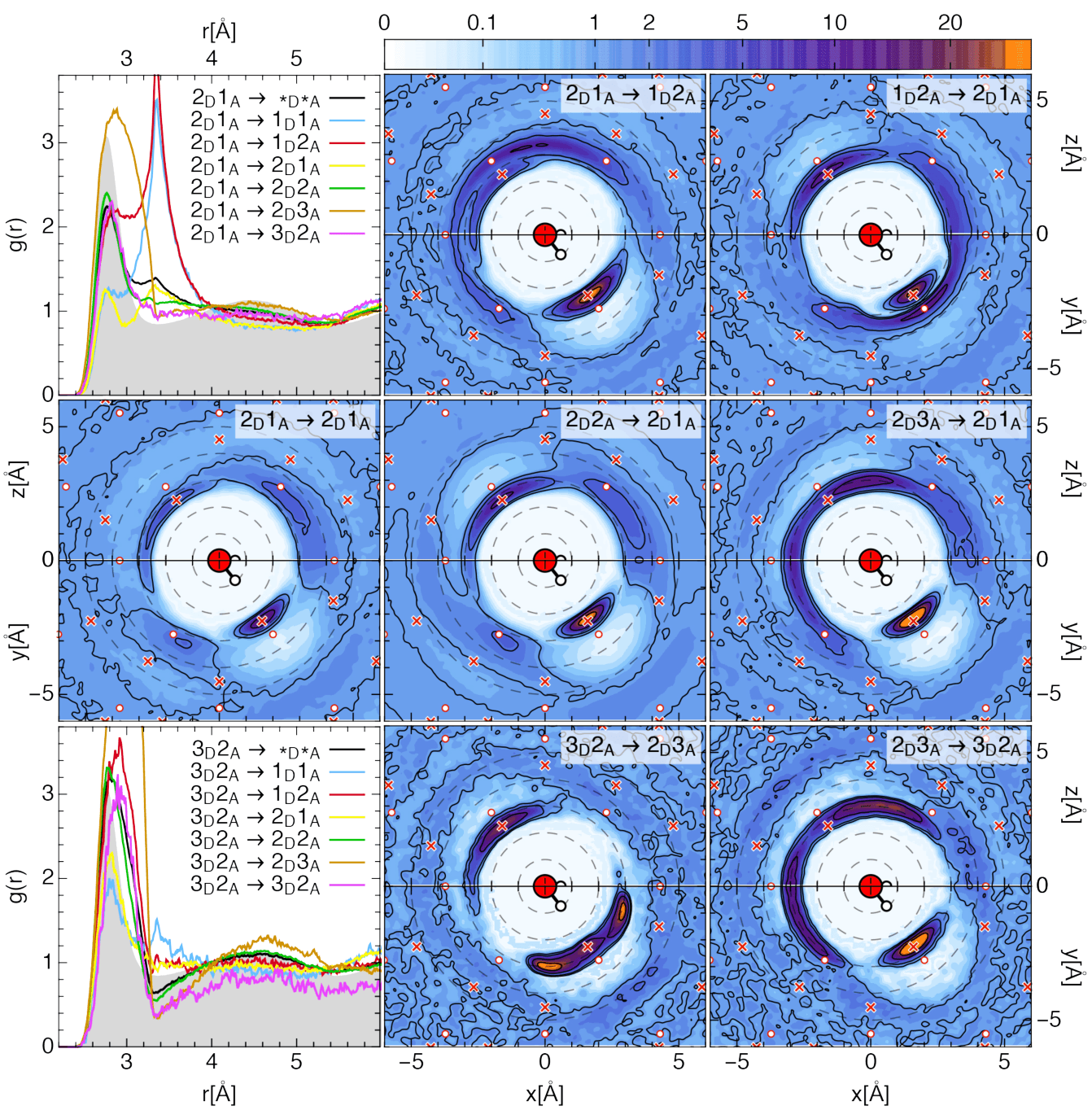}
\label{fig:OO-selected}
\end{figure*}

It is perhaps worth stressing that any decomposition of this kind 
that dissects a structural observable based on prior structural analysis, 
risks being tautological to some extent. 
For instance, the lowering of the first peak, and 
appearance of a second peak in the interstitial region 
for undercoordinated defects could be 
regarded just as hydrogen bonds caught in the act of breaking up,
rather than as a significant feature in the water network. 
An analysis of the 3D DF can identify more clearly the 
nature of the correlations between defects, telling 
apart artifacts of the analysis from genuine features
of the H-bond network.

Focusing first on the under-coordinated water environments, 
and looking in the direction of the 
donated H-bond in the 
$\da{1}{2}\rightarrow \da{\star}{\star}$ and
$\da{1}{1}\rightarrow \da{\star}{\star}$ 
correlation functions, one can see a sharp second
peak that could indeed be interpreted as arising from 
the tail of the distribution of a ``normal''
H-bond, that is identified as broken by the
PAMM fingerprint function (or the conventional
structural definition). 
On the other hand, inspection of the mirror distributions 
$\da{\star}{\star}\rightarrow \da{1}{2}$ and
$\da{\star}{\star}\rightarrow \da{1}{1}$ 
does not show a similar sharp peak just next
to the ``normal'' acceptor peak. Rather, 
it reveals the presence of strong angular 
correlations in anomalous directions,
that could be regarded as the manifestation of 
a weaker type of H-bond rather than truly 
unbound configurations or broken hydrogen bonds.
Even though all simulations in the present work
were thermostatted, and so it is not 
possible to extract rigorous dynamical information,
it is clear from inspection of the trajectories
that these weak hydrogen bonds, while
forming a smaller part of the population 
in the hydrogen bond network, are not just
fleetingly formed transition-state
structures, but rather involve meta-stable states.
It would be interesting to perform an energy
decomposition analysis as has been done in previous 
studies\cite{KuhneKhalliulin2013}, to quantify 
more precisely the enthalpic stability of these
configurations. This, however, is beyond the scope of
the current study.

Figure~\ref{fig:OO-all} quantifies the structural 
correlations for water molecules in the vicinity
of different topological defects in the H-bond network,
by fixing the coordination state of only one of the 
two oxygen atoms involved. 
Of course, one could proceed to look into pair correlation functions 
for which both species are in a prescribed coordination state.    
This more detailed analysis reveals that in many cases there appear
to be strong correlations between the position of defective coordination
states. In other terms, when fluctuations in the generally tetrahedral
network generate topological defects, such low-probability environments
appear to be clustered close to each other. 
Henchman and co-workers have performed a similar analysis
looking at the RDFs exclusively for water molecules that are
different acceptor types\cite{henc-irud10jpcb}. In particular,
they observed for example, that water molecules that were single acceptors
and triple acceptors tend to be close to each other. 
Here we considered more than 50 possible pairs of environments:
we report all the results, for different models and simulation
details, in the SI, and focus here on the most striking 
features that we could identify (Fig.~\ref{fig:OO-selected}).

Undercoordinated water molecules tend indeed to be strongly 
clustered together.  The defect-resolved RDF between $\da{2}{1}$ and
$\da{1}{2}$ environments shows a very pronounced
peak at about 3.5\AA{}. Inspection of the directionally-resolved
RDF reveals that this sharp peak is at least partly
originating from the PAMM analysis that singles out a 
hydrogen bond in the act
of breaking into a $\da{2}{1}$--$\da{1}{2}$ pair. However,
the very broad angular spread of the peak at 3.5\AA{} clearly
paints a more complicated picture in which the weakening
of the hydrogen bond is associated with greater conformational
flexibility on both the donor and the acceptor side with 
respect to a tetrahedral $\da{2}{2}$ environment
(see Fig.~\ref{fig:OO-all}). In other terms, one can see
this feature of hydrogen bonding as related to a form of
entropic stabilization. 
In addition, $\da{2}{1}$ defects are associated with unusual angular 
correlations, with two very distinct peaks that can be seen at
about 3.5\AA{} distance, well separated from typical H-bond directions.
These are seen in the $\da{2}{1}\rightarrow\da{1}{2}$,
$\da{2}{2}\rightarrow\da{1}{2}$, and also
in the $\da{2}{3}\rightarrow\da{2}{1}$ 
correlations discussed in Ref.~\cite{henc-irud10jpcb}.
 More generally, strong, anomalous angular
correlations are observed for all under-coordinated 
species (see also the whole series of defect-resolved
3D DFs in the SI), reinforcing the notion of the presence of 
a ``weak'' H-bonding mode that does not match the structural 
parameters range of a full-fledged hydrogen bond, but
that contributes to the stability of coordination defects
in liquid water.

In the liquid phase, the disorder and larger angular
flexibility of water molecules makes it difficult to
pinpoint the specific geometric features leading
to the peculiar angular correlations. We thus turned to
high density phases of water at lower temperature to
understand the origins of these correlations.
In Figure~\ref{fig:OO-selected} we overlaid to
the 3D O-O distribution functions markers that indicate 
the position of the neighboring O atoms in ice Ih (crosses)
and for a model of ice VIII \emph{expanded to match the density
of room temperature water} (circles). The 
standard hydrogen-bonded peaks correspond perfectly to 
the position of nearest neighbors in hexagonal ice, 
but the additional angular correlations found around 
undercoordinated waters are closely related to the coordination
environments in ice VIII. A relationship between ``interstitial'' 
waters and high-density phases of ice was suggested in 
Ref.~\cite{gill+13jcp} as a method to asses the accuracy of different
electronic structure methods in describing defective
environments in water. Our analysis confirms this intuition, 
and suggests that an even more representative benchmark
could be obtained by expanding the unit cell to match the
density of water at ambient conditions. 

Fig.~\ref{fig:OO-selected} also shows that 
$\da{2}{3}$ and $\da{3}{2}$ environments are very 
strongly correlated. In particular the 
$\da{3}{2}\rightarrow\da{2}{3}$ distribution function
shows two distinct peaks at the brim of the donated H-bond
region, suggesting that in most cases these configurations
are associated with bifurcated H-bonds. The inverse
distribution $\da{2}{3}\rightarrow\da{3}{2}$ 
shows the characteristic trigonal distribution of accepted H-bonds,
however with broader angular fluctuations that once again 
point at the increased flexibility associated with correlated
defective environments.

\begin{figure}
\includegraphics[width=1.0\columnwidth]{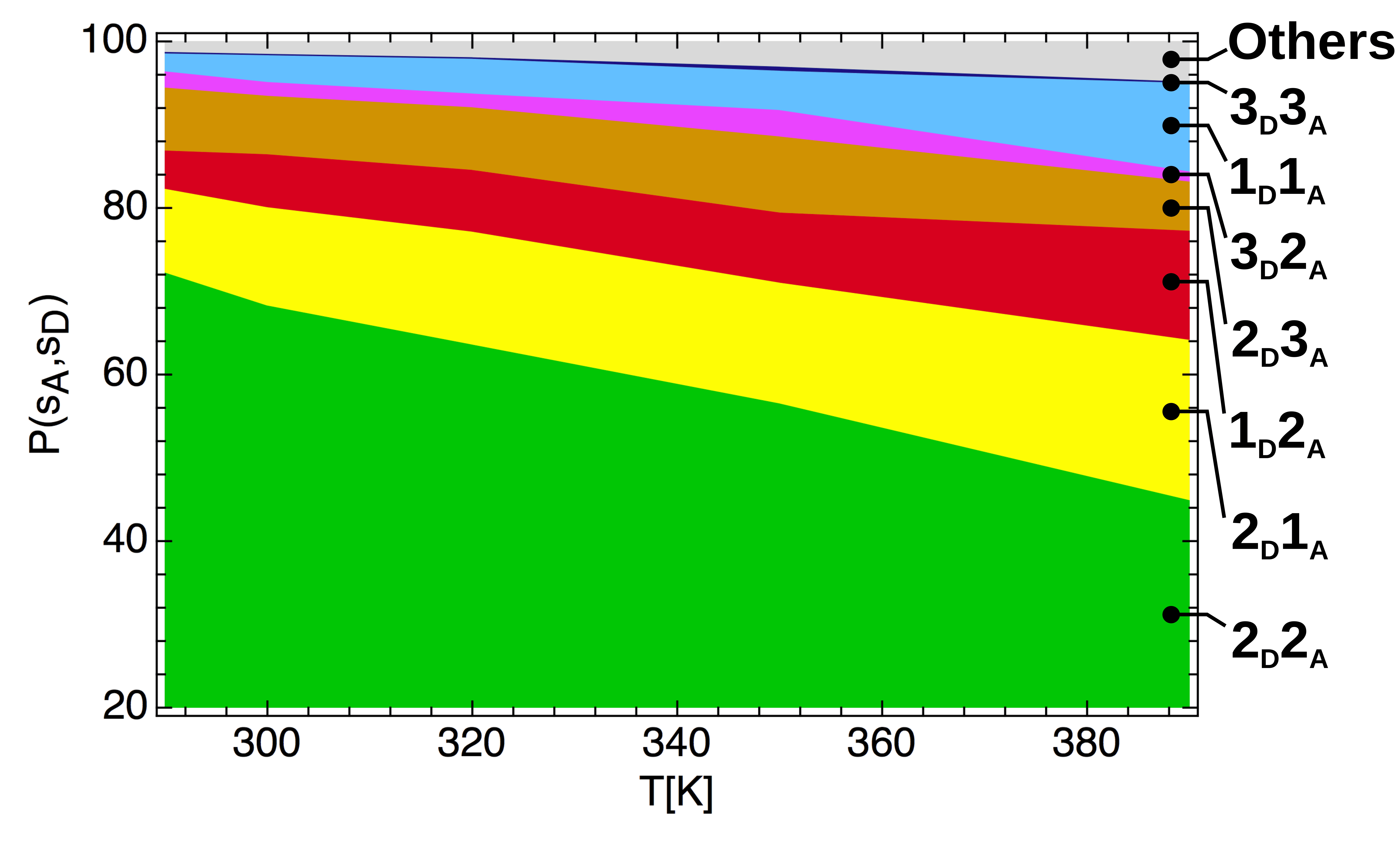}
\caption{Temperature dependence (at constant room-temperature density)
of the fraction of the main H-bonding defect states 
in the PTL simulations (BLYP+VDW, 128 water molecules).\label{fig:hb-temperature} }
\end{figure}

\section{Comparison of Water Models}
\label{sec:watermodelcomp}

In the preceding discussions we have focused on the structural
correlations of defects examining our production runs 
of the 128 water system at 300K (PTL).  In the following,
we will now describe how defect populations and correlations
change as a function of finite temperature, box size,
the quality of the electronic structure and nuclear
quantum effects. We find that topological defects
appear to be present with similar concentrations
across all the simulations. Furthermore,
the structural correlations between them also appear to be
qualitatively and sometimes even quantitatively conserved. However,
the relative concentration of different defects changes
in a subtle manner.

\begin{figure*}[hbt]
\includegraphics[width=0.7\textwidth]{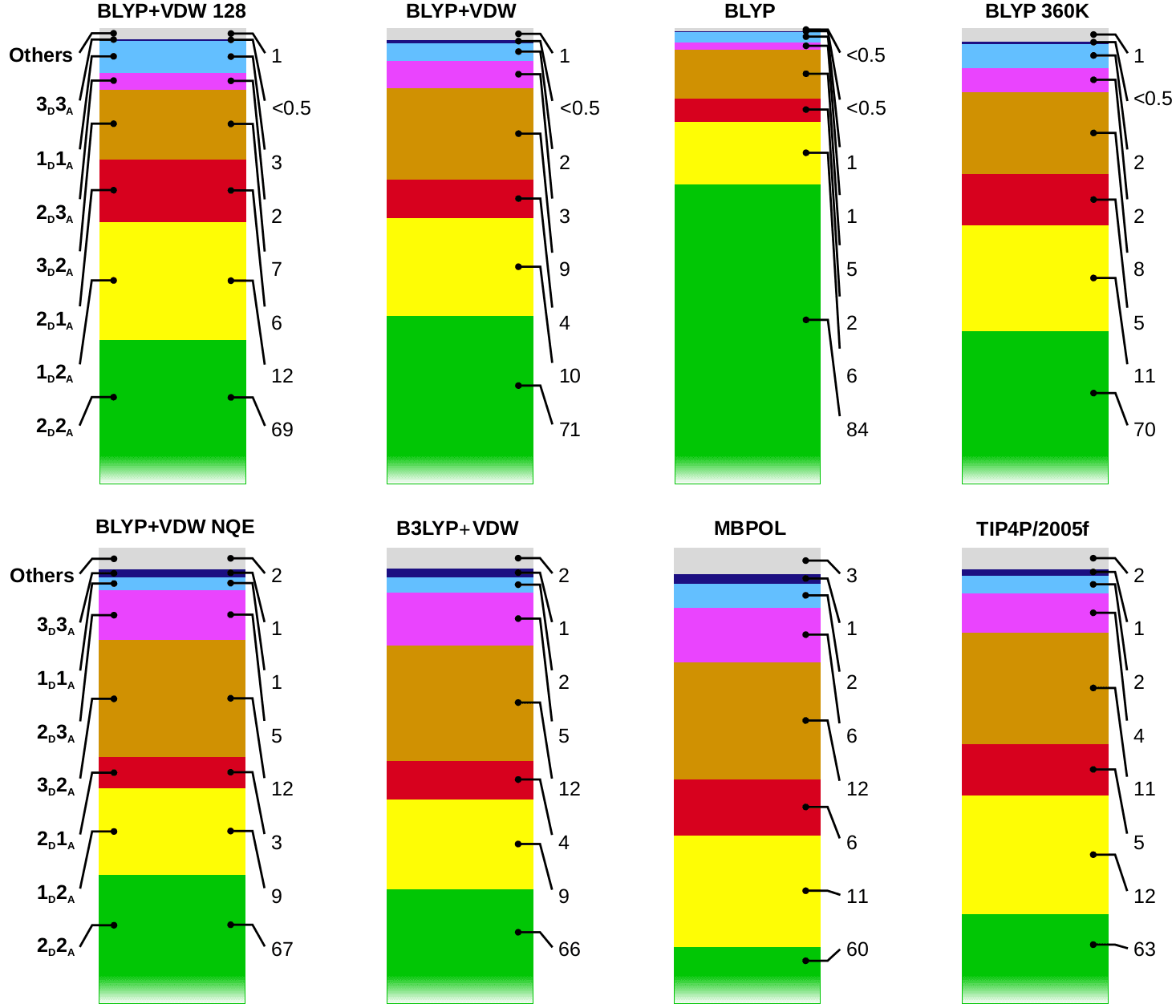}
\caption{The bar-charts show the percentage of the main defect
states considered in the text for different models explored in
this work. The label on top of each chart indicates the 
simulation protocol (functional, system size, temperature, the inclusion
of van-der-Waals corrections). On the extreme left, in bold, the
defect type  corresponding to each segment of the stack plots is indicated. 
The numbers on the right of each bar-chart indicate the 
percentage of each defect for the specified batch of simulations.
Note that the segment corresponding to the majority, tetrahedral $\da{2}{2}$ environments has been truncated for clarity.
\label{fig:model_comparison}}
\end{figure*}

\begin{figure*}[tbhp]
\includegraphics[width=0.75\textwidth]{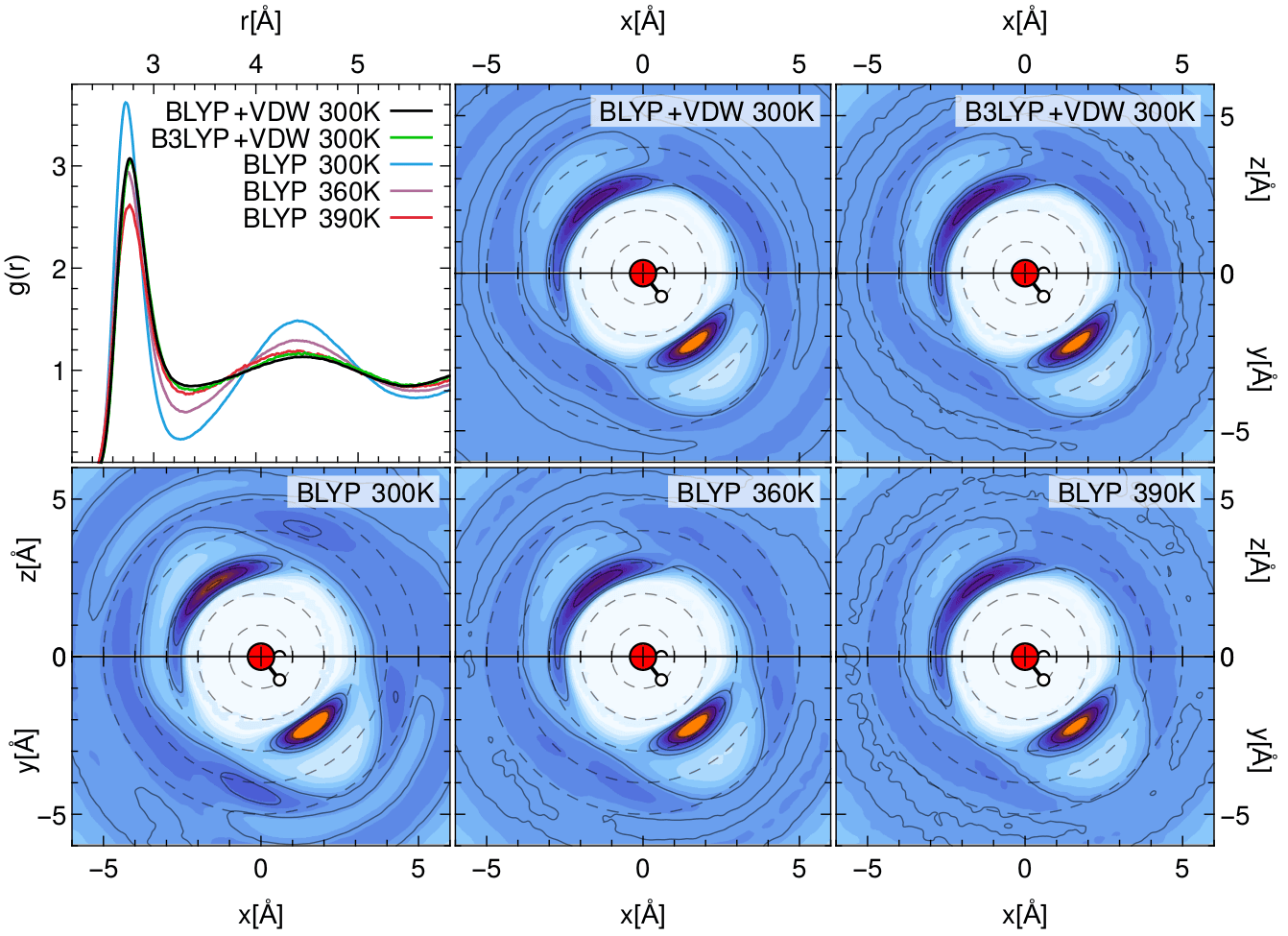}
\caption{Radial and 3D O-O distribution functions, for selected
simulation conditions. The bottom row compares BLYP-based
simulations without vdW corrections at three different
temperatures.\label{fig:blyp-comparison}}
\end{figure*}

We begin by showing how the proportion
of different types of structural defects change in water
as a function of temperature at constant, room-temperature
density. 
As one moves from 290K to 390K (Fig.~\ref{fig:hb-temperature}), 
there is a clear decrease, by about 25\%, in the
proportion of water molecules that accept and donate two 
HBs. This is in turn accompanied by an increase in
the number of structural defects.  In particular, in going from
290K to 390K there is a significant increase in the number of
water molecules accepting 2 and donating 1 HBs, accepting 1 and
donating 2 hydrogen bonds and a smaller increase in the
the number of defects accepting and donating
1 hydrogen bond. 
The concentration of overcoordinated defects such as the 
$\da{2}{3}$ and $\da{3}{2}$ stays
more or less constant, at least
in the constant density conditions we are currently simulating.
Despite the increase in the concentration of defects with
growing temperature, the 3D correlation plots reveal
the presence of well formed directed hydrogen bonds 
up to 390K (see Figure. S5-S7 in the SI).

Fig.~\ref{fig:model_comparison} shows 
the percentage of the main defects in the water network
that we obtained with various simulation protocols. 
We compare the effect of finite box size (moving from 128
to 64 water molecules, that we used for most comparisons), 
the role of nuclear quantum effects
and the use of model potentials such as TIP4P/2005
and the more recently developed MB-pol water model. 
For the most part, we see that the
relative proportion of coordination defects is not very
sensitive to the simulation protocol. In the case of MB-pol,
the number of $\da{2}{2}$ waters is lower than that of
BLYP+VDW PTWTE 64 by about 10\%, which is
compensated by a slight increase in the under-coordinated
$\da{1}{2}$ and $\da{2}{1}$ defects in the network.
It is thus rather comforting to see that all 
these various protocols for simulating liquid water 
at ambient conditions produce qualitatively consistent
results at least with respect to the types of coordination defects
in the hydrogen bond network. 
On the other hand, the predictions for physical properties
of water such as the RDF and the diffusion constant,
\emph{do} change significantly between these models.
Seeing how such changes depend on small differences in the structure and 
stability of topological defects gives some rationale for the 
difficulty in obtaining a quantitatively accurate description of 
the structural and dynamical properties of water. 

The clear outlier between the defect population plots in Fig.~\ref{fig:model_comparison} 
is the case of the BLYP functional without vdW corrections 
that leads to a dramatic increase of the fully-coordinated
tetrahedral environments. Indeed, it is well appreciated that 
standard generalized gradient approximation functionals used in AIMD
simulations lead to the overstructuring of the hydrogen bonds 
in liquid water. 
A common trick that has been suggested in early CPMD
simulations, and which has been used in many AIMD simulations thereafter,
is to increase the temperature in the simulation 
which was applied to account for the supercooled nature of DFT 
water at 300K\cite{WeiSharmaRestaGalliCar2008}. 
Typical values that have been used are
330K but there have also been suggestions that temperatures  above
400K are needed\cite{YooXantheas2011}. When this is done,
it is empirically observed that the RDF gets less structured 
and agrees more closely with experiments. 
Indeed, we see that by raising the temperature to 360K 
(the right-most stack plot in the first row of Fig.~\ref{fig:model_comparison})
one can reproduce quite accurately the defect populations 
obtained with vdW corrections.

Earlier, we alluded to the fact that one can think
of the ensemble-averaged RDF as coming from individual
contributions involving structural correlations between tetrahedral
waters together with those from the clustering of different
types of defects. To appreciate a bit better the challenge
in converging this property,
in Fig.~\ref{fig:blyp-comparison} we compare the RDFs
obtained with vdW corrections to those with the bare BLYP
functional, at 300K, 360K and 390K, to illustrate how
the choice of simulation temperature and 
the inclusion of dispersion interactions 
conspire to affect different parts of the distribution.
It is clear that van-der-Waals interactions cannot be 
fully mimicked by an increase in simulation temperature.
Without dispersion, the simulations at 360K 
reproduce the height of the first peak of the distribution,
but the long-range oscillations in the RDF are still
considerably stronger than the vdW-corrected reference. 
One has to increase the temperature up to 390K to 
approach the corrected long-range behavior, at the expense
however of lowering too much the height of the first peak.
Angular correlations, that are seen in the 3D distribution
functions, clarify the source of this discrepancy. The
bare GGA simulation shows pronounced peaks in the second
coordination shell, that correspond to the angular positions
seen in ice Ih (see also Fig~\ref{fig:OO-selected}). 
Raising the temperature broadens this peak, and shifts it 
towards the interstitial region that is characteristic
of under-coordinated defects. At the same time, the
increased temperature enhances the fluctuations and lowers
the height of the first-neighbor peak. 
There is no temperature at which thermal fluctuations 
match the effect of vdW corrections on these two components
simultaneously -- performing simulations at an artificially
increased temperature is a poor substitute for a model that
describes properly dispersion interactions. 
vdW interactions thus play an important role in tuning both
the short (first shell) and long-range (second shell
and beyond parts of the RDF although the effect is more pronounced
on the latter. Similar conclusions have also been made by
Weeks and co-workers examining classical models of
water such as SPC/E with molecular field theories\cite{Remsing2011}.

There is currently an ongoing lively debate regarding the 
level of electronic structure theory required to reproduce
structural and dynamical properties of liquid water. In particular,
several studies have advocated for the need of including a certain
amount of Hartree-Fock exchange in the exchange-correlation functional
\cite{guid+10jctc,DiStasioBiswajitZhaofengXifanCar2014}. 
For this reason, we performed AIMD simulations with the B3LYP
hybrid functional. Although we could not afford replica-exchange
simulations, we ran four independent simulations at 300K, 
for a total of more than 60ps. The 
distributions with BLYP and B3LYP (both including dispersion
interactions) show that the relative proportion of different
types of coordination states are almost quantitatively the same.
Also, radial and angular-resolved distribution functions
are remarkably similar. Thus, while it has been previously observed
that the use of hybrid functionals alone gives much better
RDFs than bare GGAs\cite{DiStasioBiswajitZhaofengXifanCar2014}, it appears that dispersion corrections 
can similarly remedy the most blatant deficiencies of BLYP,
and that combining the two does not have a major effect 
compared to applying the exchange or vdW corrections 
separately.

The delicate balance between an ice-Ih-like 
and a defective hydrogen-bond network cannot be ascribed 
to a single approximation in the electronic structure 
framework. What is more, one should not focus too much
on the RDF as the only benchmark to assess the quality
of a water model: as we have shown, angular correlations
and defect-resolved distribution functions contain much
more detailed information, and other physical properties 
(such as thermodynamic and dynamical properties~\cite{kuhn+09jctc}, or
isotope fractionation ratios~\cite{wang+14jcp})
should also be included to avoid the risk of obtaining 
an RDF that matches experiment for the wrong reasons. 
Empirical vdW corrections seem to be enough to 
reproduce the experimental RDF, but there is strong evidence
that this is largely due to a cancellation of errors between
three and four-body terms\cite{medd+15jcp}, and it has been shown
that the description of the water monomer energy is very poor
in the absence of an exact exchange correction~\cite{mora+14jctc}.

\section{Conclusions}
\label{sec:Discussion}

In this work we have used extensive AIMD simulations under a variety
of different modelling conditions to understand the complex
hydrogen bond landscape of liquid water.  We have shown that 
due to topological constraints in the hydrogen
bond network, water molecules that deviate from idealized
tetrahedral structures cluster with each other with different
propensities. In particular, we find that under-coordinated
environments display consistently strong angular correlations that 
can be traced to a weaker, but still directional, mode of the 
hydrogen bond.   
This alternative H-bonding mode, that is not recognized as such by definitions 
that are trained to identify the majority tetrahedral environment,
is related to the correlations found in a (dilated) ice VIII lattice.
Although the focus of this work has been on
structural correlations, these features have important bearing
on understanding dynamical processes in liquid water. In particular,
the clustering of these
defects suggests that the breakage and formation of hydrogen bonds
are correlated over an extended part of the network, and provides a 
mechanism to lower the barriers associated with hydrogen bond 
dynamics. In this
regard, the type of analysis we have presented here forms
a framework to rationalize, in a microscopic way, the balance
between entropic and enthalpic forces that drive fluctuations
in the hydrogen bond network.

Here, we took extra care to ensure that the AIMD simulations
we used for analysis of the HB network were sampled extensively
using ab initio replica exchange molecular dynamics. Furthermore, we
examined the sensitivity of the topological properties
of the hydrogen network to temperature, the use of dispersion corrections, 
the inclusion of exact exchange and nuclear quantum effects. Within 
the framework of ab initio methods, the inclusion of dispersion corrections
appears to have the most significant impact on the RDF compared to the 
inclusion of exact exchange or nuclear quantum effects. It is rather comforting 
to see, however, that regardless of the details of the choice of the water potential, 
the qualitative predictions for the defect populations and structures are very similar.
While there 
has been a lot of effort in trying to come up with simulation
recipes to reproduce the experimental RDF, we find that the
differences one observes from using different simulation protocols are 
quite subtle, and that for instance it is not possible to
fully mimic the effect of dispersion corrections by 
altering the temperature of the simulation. Examining
the sensitivity of defect distributions to the use of different
simulation protocols, shows that there are noticeable
differences in the relative populations of the various defects 
when varying minor computational details -- for example, from the time 
step to the basis set (see SI). It should however be 
stressed that it is very hard to assess
the statistical convergence of the minority populations, and 
that therefore we do not believe one can draw meaningful 
conclusions on the significance of these differences. 

It is clear that the radial distribution function masks a lot
of complexity in the underlying hydrogen bond network and that
the details of its structure will be
modulated by the balance in the relative 
proportions of different defects and how
they cluster with each other -- two aspects that can 
be disentangled by looking at three dimensional 
distribution functions resolved in different defect pairs. 
We believe that the analysis framework we introduce here,
based on a self-consistent, data-driven definition of the 
hydrogen bond and the study of non-trivial correlations 
between topological defects, will prove to be particularly
effective when investigating different portions
of the phase diagram of bulk water, the role of
charged defects as well as the behaviour of water 
in confinement and at interfaces.

\subsection*{Acknowledgements}
We would like to acknowledge generous allocation of CPU time by 
CSCS under the project id s466 and s553. MC and PG would like to acknowledge 
funding from CCMX and from the MPG-EPFL center for
molecular nanoscience.

See the supporting information for the details of all
of the simulations referenced in this work, a discussion of
the impact of the H-bond definition, and the complete set 
of defect population and correlation analysis.
This information is available free of charge via the 
Internet at http://pubs.acs.org

\providecommand{\latin}[1]{#1}
\providecommand*\mcitethebibliography{\thebibliography}
\csname @ifundefined\endcsname{endmcitethebibliography}
  {\let\endmcitethebibliography\endthebibliography}{}


\begin{mcitethebibliography}{61}
\providecommand*\natexlab[1]{#1}
\providecommand*\mciteSetBstSublistMode[1]{}
\providecommand*\mciteSetBstMaxWidthForm[2]{}
\providecommand*\mciteBstWouldAddEndPuncttrue
  {\def\EndOfBibitem{\unskip.}}
\providecommand*\mciteBstWouldAddEndPunctfalse
  {\let\EndOfBibitem\relax}
\providecommand*\mciteSetBstMidEndSepPunct[3]{}
\providecommand*\mciteSetBstSublistLabelBeginEnd[3]{}
\providecommand*\EndOfBibitem{}
\mciteSetBstSublistMode{f}
\mciteSetBstMaxWidthForm{subitem}{(\alph{mcitesubitemcount})}
\mciteSetBstSublistLabelBeginEnd
  {\mcitemaxwidthsubitemform\space}
  {\relax}
  {\relax}

\bibitem[Bergman(2000)]{Bergman2000}
Bergman,~D.~L. \emph{Chem. Phys.} \textbf{2000}, \emph{253}, 267 -- 282\relax
\mciteBstWouldAddEndPuncttrue
\mciteSetBstMidEndSepPunct{\mcitedefaultmidpunct}
{\mcitedefaultendpunct}{\mcitedefaultseppunct}\relax
\EndOfBibitem
\bibitem[Agmon(2012)]{AgmonReview2011}
Agmon,~N. \emph{Acc. Chem. Res.} \textbf{2012}, \emph{45}, 63--73\relax
\mciteBstWouldAddEndPuncttrue
\mciteSetBstMidEndSepPunct{\mcitedefaultmidpunct}
{\mcitedefaultendpunct}{\mcitedefaultseppunct}\relax
\EndOfBibitem
\bibitem[K{\"u}hne and Khaliullin(2013)K{\"u}hne, and
  Khaliullin]{KuhneKhalliulin2013}
K{\"u}hne,~T.~D.; Khaliullin,~R.~Z. \emph{Nat. Commun.} \textbf{2013},
  \emph{4}, 1450\relax
\mciteBstWouldAddEndPuncttrue
\mciteSetBstMidEndSepPunct{\mcitedefaultmidpunct}
{\mcitedefaultendpunct}{\mcitedefaultseppunct}\relax
\EndOfBibitem
\bibitem[Ceriotti \latin{et~al.}(2013)Ceriotti, Cuny, Parrinello, and
  Manolopoulos]{CeriottiCunyParrinelloManolopoulos2013}
Ceriotti,~M.; Cuny,~J.; Parrinello,~M.; Manolopoulos,~D.~E. \emph{Proc. Nat.
  Acad. Sci. U.S.A.} \textbf{2013}, \emph{110}, 15591--15596\relax
\mciteBstWouldAddEndPuncttrue
\mciteSetBstMidEndSepPunct{\mcitedefaultmidpunct}
{\mcitedefaultendpunct}{\mcitedefaultseppunct}\relax
\EndOfBibitem
\bibitem[Sciortino and Fornili(1989)Sciortino, and
  Fornili]{SciortinoFornili1989}
Sciortino,~F.; Fornili,~S.~L. \emph{J. Chem. Phys.} \textbf{1989}, \emph{90},
  2786--2792\relax
\mciteBstWouldAddEndPuncttrue
\mciteSetBstMidEndSepPunct{\mcitedefaultmidpunct}
{\mcitedefaultendpunct}{\mcitedefaultseppunct}\relax
\EndOfBibitem
\bibitem[Henchman and Irudayam(2010)Henchman, and Irudayam]{henc-irud10jpcb}
Henchman,~R.~H.; Irudayam,~S.~J. \emph{J. Phys. Chem. B} \textbf{2010},
  \emph{114}, 16792--16810\relax
\mciteBstWouldAddEndPuncttrue
\mciteSetBstMidEndSepPunct{\mcitedefaultmidpunct}
{\mcitedefaultendpunct}{\mcitedefaultseppunct}\relax
\EndOfBibitem
\bibitem[Kirchner \latin{et~al.}(2012)Kirchner, di~Dio, and
  Hutter]{KirchnerDioHutter2012}
Kirchner,~B.; di~Dio,~P.~J.; Hutter,~J. \emph{Multiscale Molecular Methods in
  Applied Chemistry}; Topics in Current Chemistry; Springer-Verlag Berlin,
  2012; Vol. 307; pp 109--153\relax
\mciteBstWouldAddEndPuncttrue
\mciteSetBstMidEndSepPunct{\mcitedefaultmidpunct}
{\mcitedefaultendpunct}{\mcitedefaultseppunct}\relax
\EndOfBibitem
\bibitem[Sprik \latin{et~al.}(1996)Sprik, Hutter, and Parrinello]{Sprik1996}
Sprik,~M.; Hutter,~J.; Parrinello,~M. \emph{J. Chem. Phys.} \textbf{1996},
  \emph{105}, 1142--1152\relax
\mciteBstWouldAddEndPuncttrue
\mciteSetBstMidEndSepPunct{\mcitedefaultmidpunct}
{\mcitedefaultendpunct}{\mcitedefaultseppunct}\relax
\EndOfBibitem
\bibitem[Todorova \latin{et~al.}(2006)Todorova, Seitsonen, Hutter, Kuo, and
  Mundy]{Todora2006}
Todorova,~T.; Seitsonen,~A.~P.; Hutter,~J.; Kuo,~I.-F.~W.; Mundy,~C.~J.
  \emph{J. Phys. Chem. B} \textbf{2006}, \emph{110}, 3685--3691\relax
\mciteBstWouldAddEndPuncttrue
\mciteSetBstMidEndSepPunct{\mcitedefaultmidpunct}
{\mcitedefaultendpunct}{\mcitedefaultseppunct}\relax
\EndOfBibitem
\bibitem[Guidon \latin{et~al.}(2008)Guidon, Schiffmann, Hutter, and
  VandeVondele]{GuidonSchiffmanHutterVandeVondele2008}
Guidon,~M.; Schiffmann,~F.; Hutter,~J.; VandeVondele,~J. \emph{J. Chem. Phys.}
  \textbf{2008}, \emph{128}, 214104\relax
\mciteBstWouldAddEndPuncttrue
\mciteSetBstMidEndSepPunct{\mcitedefaultmidpunct}
{\mcitedefaultendpunct}{\mcitedefaultseppunct}\relax
\EndOfBibitem
\bibitem[Silvestrelli and Parrinello(1999)Silvestrelli, and
  Parrinello]{SilvestrellliLuigiParrinello1999}
Silvestrelli,~P.~L.; Parrinello,~M. \emph{Phys. Rev. Lett.} \textbf{1999},
  \emph{82}, 3308--3311\relax
\mciteBstWouldAddEndPuncttrue
\mciteSetBstMidEndSepPunct{\mcitedefaultmidpunct}
{\mcitedefaultendpunct}{\mcitedefaultseppunct}\relax
\EndOfBibitem
\bibitem[Grossman \latin{et~al.}(2004)Grossman, Schwegler, Draeger, Gygi, and
  Galli]{GrossmanSchweglerDraegerGygiGalli2003}
Grossman,~J.~C.; Schwegler,~E.; Draeger,~E.~W.; Gygi,~F.; Galli,~G. \emph{J.
  Chem. Phys.} \textbf{2004}, \emph{120}, 300--311\relax
\mciteBstWouldAddEndPuncttrue
\mciteSetBstMidEndSepPunct{\mcitedefaultmidpunct}
{\mcitedefaultendpunct}{\mcitedefaultseppunct}\relax
\EndOfBibitem
\bibitem[Fernandez-Serra and Artacho(2004)Fernandez-Serra, and
  Artacho]{SerraArtacho2004}
Fernandez-Serra,~M.~V.; Artacho,~E. \emph{J. Chem. Phys.} \textbf{2004},
  \emph{121}, 11136--11144\relax
\mciteBstWouldAddEndPuncttrue
\mciteSetBstMidEndSepPunct{\mcitedefaultmidpunct}
{\mcitedefaultendpunct}{\mcitedefaultseppunct}\relax
\EndOfBibitem
\bibitem[Sit and Marzari(2005)Sit, and Marzari]{SitMarzari2005}
Sit,~P. H.-L.; Marzari,~N. \emph{J. Chem. Phys.} \textbf{2005}, \emph{122},
  204510\relax
\mciteBstWouldAddEndPuncttrue
\mciteSetBstMidEndSepPunct{\mcitedefaultmidpunct}
{\mcitedefaultendpunct}{\mcitedefaultseppunct}\relax
\EndOfBibitem
\bibitem[VandeVondele \latin{et~al.}(2005)VandeVondele, Mohamed, Krack, Hutter,
  Sprik, and Parrinello]{VandeVondeleMohamedKrackHutterSprikParrinello2005}
VandeVondele,~J.; Mohamed,~F.; Krack,~M.; Hutter,~J.; Sprik,~M.; Parrinello,~M.
  \emph{J. Chem. Phys.} \textbf{2005}, \emph{122}, 014515\relax
\mciteBstWouldAddEndPuncttrue
\mciteSetBstMidEndSepPunct{\mcitedefaultmidpunct}
{\mcitedefaultendpunct}{\mcitedefaultseppunct}\relax
\EndOfBibitem
\bibitem[Lee and Tuckerman(2006)Lee, and Tuckerman]{LeeTuckerman2006b}
Lee,~H.-S.; Tuckerman,~M.~E. \emph{J. Chem. Phys.} \textbf{2006}, \emph{125},
  154507\relax
\mciteBstWouldAddEndPuncttrue
\mciteSetBstMidEndSepPunct{\mcitedefaultmidpunct}
{\mcitedefaultendpunct}{\mcitedefaultseppunct}\relax
\EndOfBibitem
\bibitem[Kuo \latin{et~al.}(2004)Kuo, Mundy, McGrath, Siepmann, VandeVondele,
  Sprik, Hutter, Chen, Klein, Mohamed, Krack, and
  Parrinello]{KuoMundyMcGrathSiepmannVandeVondeleSprikHutterKleinMohamedKrackParrinello2004}
Kuo,~I.-F.~W.; Mundy,~C.~J.; McGrath,~M.~J.; Siepmann,~J.~I.; VandeVondele,~J.;
  Sprik,~M.; Hutter,~J.; Chen,~B.; Klein,~M.~L.; Mohamed,~F.; Krack,~M.;
  Parrinello,~M. \emph{J. Phys. Chem. B} \textbf{2004}, \emph{108},
  12990--12998\relax
\mciteBstWouldAddEndPuncttrue
\mciteSetBstMidEndSepPunct{\mcitedefaultmidpunct}
{\mcitedefaultendpunct}{\mcitedefaultseppunct}\relax
\EndOfBibitem
\bibitem[Zhang \latin{et~al.}(2011)Zhang, Donadio, Gygi, and Galli]{Zhang2011}
Zhang,~C.; Donadio,~D.; Gygi,~F.; Galli,~G. \emph{J. Chem. Theo. Comput.}
  \textbf{2011}, \emph{7}, 1443--1449\relax
\mciteBstWouldAddEndPuncttrue
\mciteSetBstMidEndSepPunct{\mcitedefaultmidpunct}
{\mcitedefaultendpunct}{\mcitedefaultseppunct}\relax
\EndOfBibitem
\bibitem[Lin \latin{et~al.}(2009)Lin, Seitsonen, Coutinho-Neto, Tavernelli, and
  Rothlisberger]{LinSeitsonenCoutinhoMauricioTavernelliRothlisberger2009}
Lin,~I.-C.; Seitsonen,~A.~P.; Coutinho-Neto,~M.~D.; Tavernelli,~I.;
  Rothlisberger,~U. \emph{J. Phys. Chem. B} \textbf{2009}, \emph{113},
  1127--1131\relax
\mciteBstWouldAddEndPuncttrue
\mciteSetBstMidEndSepPunct{\mcitedefaultmidpunct}
{\mcitedefaultendpunct}{\mcitedefaultseppunct}\relax
\EndOfBibitem
\bibitem[Santra \latin{et~al.}(2009)Santra, Michaelides, and
  Scheffler]{SantraMichaelidesScheffler2009}
Santra,~B.; Michaelides,~A.; Scheffler,~M. \emph{J. Chem. Phys.} \textbf{2009},
  \emph{131}\relax
\mciteBstWouldAddEndPuncttrue
\mciteSetBstMidEndSepPunct{\mcitedefaultmidpunct}
{\mcitedefaultendpunct}{\mcitedefaultseppunct}\relax
\EndOfBibitem
\bibitem[Jonchiere \latin{et~al.}(2011)Jonchiere, Seitsonen, Ferlat, Saitta,
  and Vuilleumier]{JohnchiereSeitsonenFerlatSaittaVuilleumier2011}
Jonchiere,~R.; Seitsonen,~A.~P.; Ferlat,~G.; Saitta,~A.~M.; Vuilleumier,~R.
  \emph{J. Chem. Phys.} \textbf{2011}, \emph{135}, 154503\relax
\mciteBstWouldAddEndPuncttrue
\mciteSetBstMidEndSepPunct{\mcitedefaultmidpunct}
{\mcitedefaultendpunct}{\mcitedefaultseppunct}\relax
\EndOfBibitem
\bibitem[Zhang \latin{et~al.}(2011)Zhang, Wu, Galli, and
  Gygi]{CuiWuGalliGygi2011}
Zhang,~C.; Wu,~J.; Galli,~G.; Gygi,~F. \emph{J. Chem. Theo. Comput.}
  \textbf{2011}, \emph{7}, 3054--3061\relax
\mciteBstWouldAddEndPuncttrue
\mciteSetBstMidEndSepPunct{\mcitedefaultmidpunct}
{\mcitedefaultendpunct}{\mcitedefaultseppunct}\relax
\EndOfBibitem
\bibitem[M{\o}gelh{\o}j \latin{et~al.}(2011)M{\o}gelh{\o}j, Kelkkanen,
  Wikfeldt, Schi{\o}tz, Mortensen, Pettersson, Lundqvist, Jacobsen, Nilsson,
  and
  N{\o}rskov]{AndreasKelkkanenAndreWikfeldtJakobJorgenLArsJacobsenNillssonJens2011}
M{\o}gelh{\o}j,~A.; Kelkkanen,~A.~K.; Wikfeldt,~K.~T.; Schi{\o}tz,~J.;
  Mortensen,~J.~J.; Pettersson,~L. G.~M.; Lundqvist,~B.~I.; Jacobsen,~K.~W.;
  Nilsson,~A.; N{\o}rskov,~J.~K. \emph{J. Phys. Chem. B} \textbf{2011},
  \emph{115}, 14149--14160\relax
\mciteBstWouldAddEndPuncttrue
\mciteSetBstMidEndSepPunct{\mcitedefaultmidpunct}
{\mcitedefaultendpunct}{\mcitedefaultseppunct}\relax
\EndOfBibitem
\bibitem[Chen \latin{et~al.}(2003)Chen, Ivanov, Klein, and
  Parrinello]{Chen2003}
Chen,~B.; Ivanov,~I.; Klein,~M.~L.; Parrinello,~M. \emph{Phys. Rev. Lett.}
  \textbf{2003}, \emph{91}, 215503\relax
\mciteBstWouldAddEndPuncttrue
\mciteSetBstMidEndSepPunct{\mcitedefaultmidpunct}
{\mcitedefaultendpunct}{\mcitedefaultseppunct}\relax
\EndOfBibitem
\bibitem[Morrone and Car(2008)Morrone, and Car]{MorroneCar2008}
Morrone,~J.~A.; Car,~R. \emph{Phys. Rev. Lett.} \textbf{2008}, \emph{101},
  017801\relax
\mciteBstWouldAddEndPuncttrue
\mciteSetBstMidEndSepPunct{\mcitedefaultmidpunct}
{\mcitedefaultendpunct}{\mcitedefaultseppunct}\relax
\EndOfBibitem
\bibitem[{Del Ben} \latin{et~al.}(2013){Del Ben}, Sch{\"{o}}nherr, Hutter, and
  Vandevondele]{delb+13jpcl}
{Del Ben},~M.; Sch{\"{o}}nherr,~M.; Hutter,~J.; Vandevondele,~J. \emph{J. Phys.
  Chem. Lett.} \textbf{2013}, \emph{4}, 3753--3759\relax
\mciteBstWouldAddEndPuncttrue
\mciteSetBstMidEndSepPunct{\mcitedefaultmidpunct}
{\mcitedefaultendpunct}{\mcitedefaultseppunct}\relax
\EndOfBibitem
\bibitem[DiStasio \latin{et~al.}(2014)DiStasio, Santra, Li, Wu, and
  Car]{DiStasioBiswajitZhaofengXifanCar2014}
DiStasio,~R.~A.; Santra,~B.; Li,~Z.; Wu,~X.; Car,~R. \emph{J. Chem. Phys.}
  \textbf{2014}, \emph{141}, --\relax
\mciteBstWouldAddEndPuncttrue
\mciteSetBstMidEndSepPunct{\mcitedefaultmidpunct}
{\mcitedefaultendpunct}{\mcitedefaultseppunct}\relax
\EndOfBibitem
\bibitem[Bonomi and Parrinello(2010)Bonomi, and Parrinello]{bono-parr10prl}
Bonomi,~M.; Parrinello,~M. \emph{Phys. Rev. Lett.} \textbf{2010}, \emph{104},
  190601\relax
\mciteBstWouldAddEndPuncttrue
\mciteSetBstMidEndSepPunct{\mcitedefaultmidpunct}
{\mcitedefaultendpunct}{\mcitedefaultseppunct}\relax
\EndOfBibitem
\bibitem[Gasparotto and Ceriotti(2014)Gasparotto, and Ceriotti]{gasp-ceri14jcp}
Gasparotto,~P.; Ceriotti,~M. \emph{J. Chem. Phys.} \textbf{2014}, \emph{141},
  174110\relax
\mciteBstWouldAddEndPuncttrue
\mciteSetBstMidEndSepPunct{\mcitedefaultmidpunct}
{\mcitedefaultendpunct}{\mcitedefaultseppunct}\relax
\EndOfBibitem
\bibitem[Yoo and Xantheas(2011)Yoo, and Xantheas]{YooXantheas2011}
Yoo,~S.; Xantheas,~S.~S. \emph{J. Chem. Phys.} \textbf{2011}, \emph{134},
  --\relax
\mciteBstWouldAddEndPuncttrue
\mciteSetBstMidEndSepPunct{\mcitedefaultmidpunct}
{\mcitedefaultendpunct}{\mcitedefaultseppunct}\relax
\EndOfBibitem
\bibitem[Akin-Ojo and Wang(2011)Akin-Ojo, and Wang]{AkinOjoWang2011}
Akin-Ojo,~O.; Wang,~F. \emph{Chem. Phys. Lett.} \textbf{2011}, \emph{513}, 59
  -- 62\relax
\mciteBstWouldAddEndPuncttrue
\mciteSetBstMidEndSepPunct{\mcitedefaultmidpunct}
{\mcitedefaultendpunct}{\mcitedefaultseppunct}\relax
\EndOfBibitem
\bibitem[Stanley and Teixeira(1980)Stanley, and Teixeira]{StanleyTeixeira1980}
Stanley,~H.~E.; Teixeira,~J. \emph{J. Chem. Phys.} \textbf{1980}, \emph{73},
  3404--3422\relax
\mciteBstWouldAddEndPuncttrue
\mciteSetBstMidEndSepPunct{\mcitedefaultmidpunct}
{\mcitedefaultendpunct}{\mcitedefaultseppunct}\relax
\EndOfBibitem
\bibitem[Soper and Phillips(1986)Soper, and Phillips]{Soper1986}
Soper,~A.~K.; Phillips,~M.~G. \emph{Chem. Phys.} \textbf{1986}, \emph{107},
  47--60\relax
\mciteBstWouldAddEndPuncttrue
\mciteSetBstMidEndSepPunct{\mcitedefaultmidpunct}
{\mcitedefaultendpunct}{\mcitedefaultseppunct}\relax
\EndOfBibitem
\bibitem[Hassanali \latin{et~al.}(2013)Hassanali, Giberti, Cuny, Kühne, and
  Parrinello]{HassanaliGibertiCunyKuhneParrinello2013}
Hassanali,~A.; Giberti,~F.; Cuny,~J.; Kühne,~T.~D.; Parrinello,~M. \emph{Proc.
  Nat. Acad. Sci. U.S.A.} \textbf{2013}, 13723\relax
\mciteBstWouldAddEndPuncttrue
\mciteSetBstMidEndSepPunct{\mcitedefaultmidpunct}
{\mcitedefaultendpunct}{\mcitedefaultseppunct}\relax
\EndOfBibitem
\bibitem[VandeVondele \latin{et~al.}(2005)VandeVondele, Krack, Mohamed,
  Parrinello, Chassaing, and
  Hutter]{VandeVondeleKrackMohamedParrinelloChassaingHutter2005}
VandeVondele,~J.; Krack,~M.; Mohamed,~F.; Parrinello,~M.; Chassaing,~T.;
  Hutter,~J. \emph{Comput. Phys. Commun.} \textbf{2005}, \emph{167},
  103--128\relax
\mciteBstWouldAddEndPuncttrue
\mciteSetBstMidEndSepPunct{\mcitedefaultmidpunct}
{\mcitedefaultendpunct}{\mcitedefaultseppunct}\relax
\EndOfBibitem
\bibitem[Ceriotti \latin{et~al.}(2014)Ceriotti, More, and
  Manolopoulos]{ceri+14cpc}
Ceriotti,~M.; More,~J.; Manolopoulos,~D.~E. \emph{Comput. Phys. Commun.}
  \textbf{2014}, \emph{185}, 1019--1026\relax
\mciteBstWouldAddEndPuncttrue
\mciteSetBstMidEndSepPunct{\mcitedefaultmidpunct}
{\mcitedefaultendpunct}{\mcitedefaultseppunct}\relax
\EndOfBibitem
\bibitem[Grimme \latin{et~al.}(2010)Grimme, Antony, Ehrlich, and
  Krieg]{grim+10jcp}
Grimme,~S.; Antony,~J.; Ehrlich,~S.; Krieg,~H. \emph{J. Chem. Phys.}
  \textbf{2010}, \emph{132}, 154104\relax
\mciteBstWouldAddEndPuncttrue
\mciteSetBstMidEndSepPunct{\mcitedefaultmidpunct}
{\mcitedefaultendpunct}{\mcitedefaultseppunct}\relax
\EndOfBibitem
\bibitem[Lee \latin{et~al.}(1988)Lee, Yang, and Parr]{LeeYangParr1988}
Lee,~C.; Yang,~W.; Parr,~R.~G. \emph{Phys. Rev.} \textbf{1988}, \emph{B37},
  785--789\relax
\mciteBstWouldAddEndPuncttrue
\mciteSetBstMidEndSepPunct{\mcitedefaultmidpunct}
{\mcitedefaultendpunct}{\mcitedefaultseppunct}\relax
\EndOfBibitem
\bibitem[Goedecker \latin{et~al.}(1996)Goedecker, Teter, and
  Hutter]{GoedeckerTeterHutter1996}
Goedecker,~S.; Teter,~M.; Hutter,~J. \emph{Phys. Rev.} \textbf{1996},
  \emph{B54}, 1703--1710\relax
\mciteBstWouldAddEndPuncttrue
\mciteSetBstMidEndSepPunct{\mcitedefaultmidpunct}
{\mcitedefaultendpunct}{\mcitedefaultseppunct}\relax
\EndOfBibitem
\bibitem[Bussi \latin{et~al.}(2007)Bussi, Donadio, and Parrinello]{buss+07jcp}
Bussi,~G.; Donadio,~D.; Parrinello,~M. \emph{J. Chem. Phys.} \textbf{2007},
  \emph{126}, 14101\relax
\mciteBstWouldAddEndPuncttrue
\mciteSetBstMidEndSepPunct{\mcitedefaultmidpunct}
{\mcitedefaultendpunct}{\mcitedefaultseppunct}\relax
\EndOfBibitem
\bibitem[Ceriotti \latin{et~al.}(2009)Ceriotti, Bussi, and
  Parrinello]{ceri+09prl}
Ceriotti,~M.; Bussi,~G.; Parrinello,~M. \emph{Phys. Rev. Lett.} \textbf{2009},
  \emph{102}, 020601\relax
\mciteBstWouldAddEndPuncttrue
\mciteSetBstMidEndSepPunct{\mcitedefaultmidpunct}
{\mcitedefaultendpunct}{\mcitedefaultseppunct}\relax
\EndOfBibitem
\bibitem[Ceriotti \latin{et~al.}(2010)Ceriotti, Bussi, and
  Parrinello]{ceri+10jctc}
Ceriotti,~M.; Bussi,~G.; Parrinello,~M. \emph{J. Chem. Theo. Comput.}
  \textbf{2010}, \emph{6}, 1170--1180\relax
\mciteBstWouldAddEndPuncttrue
\mciteSetBstMidEndSepPunct{\mcitedefaultmidpunct}
{\mcitedefaultendpunct}{\mcitedefaultseppunct}\relax
\EndOfBibitem
\bibitem[Ceriotti and Manolopoulos(2012)Ceriotti, and
  Manolopoulos]{ceri-mano12prl}
Ceriotti,~M.; Manolopoulos,~D.~E. \emph{Phys. Rev. Lett.} \textbf{2012},
  \emph{109}, 100604\relax
\mciteBstWouldAddEndPuncttrue
\mciteSetBstMidEndSepPunct{\mcitedefaultmidpunct}
{\mcitedefaultendpunct}{\mcitedefaultseppunct}\relax
\EndOfBibitem
\bibitem[Guidon \latin{et~al.}(2010)Guidon, Hutter, and
  VandeVondele]{guid+10jctc}
Guidon,~M.; Hutter,~J.; VandeVondele,~J. \emph{J. Chem. Theo. Comput.}
  \textbf{2010}, \emph{6}, 2348--2364\relax
\mciteBstWouldAddEndPuncttrue
\mciteSetBstMidEndSepPunct{\mcitedefaultmidpunct}
{\mcitedefaultendpunct}{\mcitedefaultseppunct}\relax
\EndOfBibitem
\bibitem[Gonz{\'{a}}lez and Abascal(2011)Gonz{\'{a}}lez, and
  Abascal]{gonz-abas11jcp}
Gonz{\'{a}}lez,~M.~a.; Abascal,~J. L.~F. \emph{J. Chem. Phys.} \textbf{2011},
  \emph{135}, 224516\relax
\mciteBstWouldAddEndPuncttrue
\mciteSetBstMidEndSepPunct{\mcitedefaultmidpunct}
{\mcitedefaultendpunct}{\mcitedefaultseppunct}\relax
\EndOfBibitem
\bibitem[Babin \latin{et~al.}(2013)Babin, Leforestier, and Paesani]{MBPOL1}
Babin,~V.; Leforestier,~C.; Paesani,~F. \emph{J. Chem. Theo. Comput.}
  \textbf{2013}, \emph{9}, 5395--5403\relax
\mciteBstWouldAddEndPuncttrue
\mciteSetBstMidEndSepPunct{\mcitedefaultmidpunct}
{\mcitedefaultendpunct}{\mcitedefaultseppunct}\relax
\EndOfBibitem
\bibitem[Babin \latin{et~al.}(2014)Babin, Medders, and Paesani]{MBPOL2}
Babin,~V.; Medders,~G.~R.; Paesani,~F. \emph{J. Chem. Theo. Comput.}
  \textbf{2014}, \emph{10}, 1599--1607\relax
\mciteBstWouldAddEndPuncttrue
\mciteSetBstMidEndSepPunct{\mcitedefaultmidpunct}
{\mcitedefaultendpunct}{\mcitedefaultseppunct}\relax
\EndOfBibitem
\bibitem[Medders \latin{et~al.}(2014)Medders, Babin, and Paesani]{MBPOL3}
Medders,~G.~R.; Babin,~V.; Paesani,~F. \emph{J. Chem. Theo. Comput.}
  \textbf{2014}, \emph{10}, 2906--2910\relax
\mciteBstWouldAddEndPuncttrue
\mciteSetBstMidEndSepPunct{\mcitedefaultmidpunct}
{\mcitedefaultendpunct}{\mcitedefaultseppunct}\relax
\EndOfBibitem
\bibitem[Kumar \latin{et~al.}(2007)Kumar, Schmidt, and Skinner]{kuma+07jcp}
Kumar,~R.; Schmidt,~J.~R.; Skinner,~J.~L. \emph{J. Chem. Phys.} \textbf{2007},
  \emph{126}, 204107\relax
\mciteBstWouldAddEndPuncttrue
\mciteSetBstMidEndSepPunct{\mcitedefaultmidpunct}
{\mcitedefaultendpunct}{\mcitedefaultseppunct}\relax
\EndOfBibitem
\bibitem[Laage and Hynes(2008)Laage, and Hynes]{LaageHynes2008}
Laage,~D.; Hynes,~J.~T. \emph{J. Phys. Chem. B} \textbf{2008}, \emph{112},
  14230--14242\relax
\mciteBstWouldAddEndPuncttrue
\mciteSetBstMidEndSepPunct{\mcitedefaultmidpunct}
{\mcitedefaultendpunct}{\mcitedefaultseppunct}\relax
\EndOfBibitem
\bibitem[Lawrence and Skinner(2003)Lawrence, and Skinner]{lawr-skin03cpl}
Lawrence,~C.; Skinner,~J. \emph{Chem. Phys. Lett.} \textbf{2003}, \emph{369},
  472--477\relax
\mciteBstWouldAddEndPuncttrue
\mciteSetBstMidEndSepPunct{\mcitedefaultmidpunct}
{\mcitedefaultendpunct}{\mcitedefaultseppunct}\relax
\EndOfBibitem
\bibitem[Auer \latin{et~al.}(2007)Auer, Kumar, Schmidt, and
  Skinner]{auer+07pnas}
Auer,~B.; Kumar,~R.; Schmidt,~J.~R.; Skinner,~J.~L. \emph{Proc. Natl. Acad.
  Sci. USA} \textbf{2007}, \emph{104}, 14215--14220\relax
\mciteBstWouldAddEndPuncttrue
\mciteSetBstMidEndSepPunct{\mcitedefaultmidpunct}
{\mcitedefaultendpunct}{\mcitedefaultseppunct}\relax
\EndOfBibitem
\bibitem[Wikfeldt \latin{et~al.}(2011)Wikfeldt, Nilsson, and
  Pettersson]{WikfeldtNilssonPettersson2011}
Wikfeldt,~K.~T.; Nilsson,~A.; Pettersson,~L. G.~M. \emph{Phys. Chem. Chem.
  Phys.} \textbf{2011}, \emph{13}, 19918--19924\relax
\mciteBstWouldAddEndPuncttrue
\mciteSetBstMidEndSepPunct{\mcitedefaultmidpunct}
{\mcitedefaultendpunct}{\mcitedefaultseppunct}\relax
\EndOfBibitem
\bibitem[Gillan \latin{et~al.}(2013)Gillan, Alf{\`{e}}, Bart{\'{o}}k, and
  Cs{\'{a}}nyi]{gill+13jcp}
Gillan,~M.~J.; Alf{\`{e}},~D.; Bart{\'{o}}k,~a.~P.; Cs{\'{a}}nyi,~G. \emph{J.
  Chem. Phys.} \textbf{2013}, \emph{139}, 244504\relax
\mciteBstWouldAddEndPuncttrue
\mciteSetBstMidEndSepPunct{\mcitedefaultmidpunct}
{\mcitedefaultendpunct}{\mcitedefaultseppunct}\relax
\EndOfBibitem
\bibitem[Chen \latin{et~al.}(2008)Chen, Sharma, Resta, Galli, and
  Car]{WeiSharmaRestaGalliCar2008}
Chen,~W.; Sharma,~M.; Resta,~R.; Galli,~G.; Car,~R. \emph{Phys. Rev. B:
  Condens. Matter Mater. Phys.} \textbf{2008}, \emph{77}, 245114\relax
\mciteBstWouldAddEndPuncttrue
\mciteSetBstMidEndSepPunct{\mcitedefaultmidpunct}
{\mcitedefaultendpunct}{\mcitedefaultseppunct}\relax
\EndOfBibitem
\bibitem[Remsing \latin{et~al.}(2011)Remsing, Rodgers, and Weeks]{Remsing2011}
Remsing,~R.~C.; Rodgers,~J.~M.; Weeks,~J.~D. \emph{Journal of Statistical
  Physics} \textbf{2011}, \emph{145}, 313--334\relax
\mciteBstWouldAddEndPuncttrue
\mciteSetBstMidEndSepPunct{\mcitedefaultmidpunct}
{\mcitedefaultendpunct}{\mcitedefaultseppunct}\relax
\EndOfBibitem
\bibitem[K{\"{u}}hne \latin{et~al.}(2009)K{\"{u}}hne, Krack, and
  Parrinello]{kuhn+09jctc}
K{\"{u}}hne,~T.~D.; Krack,~M.; Parrinello,~M. \emph{J. Chem. Theo. Comput.}
  \textbf{2009}, \emph{5}, 235--241\relax
\mciteBstWouldAddEndPuncttrue
\mciteSetBstMidEndSepPunct{\mcitedefaultmidpunct}
{\mcitedefaultendpunct}{\mcitedefaultseppunct}\relax
\EndOfBibitem
\bibitem[Wang \latin{et~al.}(2014)Wang, Ceriotti, and Markland]{wang+14jcp}
Wang,~L.; Ceriotti,~M.; Markland,~T.~E. \emph{J. Chem. Phys.} \textbf{2014},
  \emph{141}, 104502\relax
\mciteBstWouldAddEndPuncttrue
\mciteSetBstMidEndSepPunct{\mcitedefaultmidpunct}
{\mcitedefaultendpunct}{\mcitedefaultseppunct}\relax
\EndOfBibitem
\bibitem[Medders \latin{et~al.}(2015)Medders, G{\"{o}}tz, Morales, Bajaj, and
  Paesani]{medd+15jcp}
Medders,~G.~R.; G{\"{o}}tz,~A.~W.; Morales,~M.~A.; Bajaj,~P.; Paesani,~F.
  \emph{J. Chem. Phys.} \textbf{2015}, \emph{143}, 104102\relax
\mciteBstWouldAddEndPuncttrue
\mciteSetBstMidEndSepPunct{\mcitedefaultmidpunct}
{\mcitedefaultendpunct}{\mcitedefaultseppunct}\relax
\EndOfBibitem
\bibitem[Morales \latin{et~al.}(2014)Morales, Gergely, McMinis, McMahon, Kim,
  and Ceperley]{mora+14jctc}
Morales,~M.~a.; Gergely,~J.~R.; McMinis,~J.; McMahon,~J.~M.; Kim,~J.;
  Ceperley,~D.~M. \emph{J. Chem. Theo. Comput.} \textbf{2014}, \emph{10},
  2355--2362\relax
\mciteBstWouldAddEndPuncttrue
\mciteSetBstMidEndSepPunct{\mcitedefaultmidpunct}
{\mcitedefaultendpunct}{\mcitedefaultseppunct}\relax
\EndOfBibitem
\end{mcitethebibliography}
\end{document}